\documentclass[prb, 11pt, aps, showkeys, preprintnumbers,amsmath,amssymb, longbibliography]{revtex4-1}

\usepackage{graphicx}
\usepackage{dcolumn}
\usepackage[bookmarks, colorlinks=true, breaklinks]{hyperref}
\hypersetup{linkcolor=blue,citecolor=blue,filecolor=black,urlcolor=blue}
\usepackage{color} 

\usepackage{url,hyperref,lineno,microtype}
\usepackage{bm}
\newcommand{\bra}{\left< }
\newcommand{\ket}{\right>}
\newcommand{\ovl}[1]{\overline{#1}}

\begin{document}

\title{Negative thermal expansion near the precipice of structural stability in open perovskites} 

%\author{Connor A. Occhialini\,$^{1,2}$\footnote{Present Affiliation: Department of Physics, Massachusetts Institute of Technology, Cambridge, Massachusetts 02139, USA }, Gian G. Guzm\'{a}n-Verri\,$^{3,4}$\footnote{email: gian.guzman@ucr.ac.cr}, Sahan U. Handunkanda$^{1,2}$\, and Jason N. Hancock\,$^{1,2}$\footnote{email: jason.hancock@uconn.edu}}
%\affiliation{$^{1}$Department of Physics, University of Connecticut, Storrs, Connecticut 06269, USA,}
%\affiliation{$^{2}$ Institute of Materials Science, University of Connecticut, Storrs, Connecticut 06269, USA }
%\affiliation{$^{3}$ Centro de Investigaci\'{o}n en Ciencia e Ingenier\'{i}a de Materiales (CICIMA), Universidad de Costa Rica, San Jos\'{e}, Costa Rica 11501}
%\affiliation{$^{4}$ Materials Science Division, Argonne National Laboratory, Argonne, Illinois 60439, USA}

\author{Connor A. Occhialini\,$^{1,2}$\footnote{Present affiliation: Department of Physics, Massachusetts Institute of Technology, Cambridge, MA 02139, USA }, Gian G. Guzm\'{a}n-Verri\,$^{3,4}$\footnote{Email: gian.guzman@ucr.ac.cr}, Sahan U. Handunkanda$^{1,2}$\, and Jason N. Hancock\,$^{1,2}$\footnote{Email: jason.hancock@uconn.edu}}
\affiliation{$^{1}$Department of Physics, University of Connecticut, Storrs, Connecticut 06269, USA,}
\affiliation{$^{2}$ Institute of Materials Science, University of Connecticut, Storrs, Connecticut 06269, USA }
\affiliation{$^{3}$ Centro de Investigaci\'{o}n en Ciencia e Ingenier\'{i}a de Materiales (CICIMA), Universidad de Costa Rica, San Jos\'{e}, Costa Rica 11501}
\affiliation{$^{4}$ Materials Science Division, Argonne National Laboratory, Argonne, Illinois 60439, USA}

\date{\today}


\begin{abstract}

Negative thermal expansion (NTE) describes the anomalous propensity of materials to shrink when heated.  Since its discovery, the NTE effect has been found in a wide variety of materials with an array of magnetic, electronic and structural properties. In some cases, the NTE originates from phase competition arising from the electronic or magnetic degrees of freedom but we here focus on a particular class of NTE which originates from intrinsic dynamical origins related to the lattice degrees of freedom, a property we term \textit{structural} negative thermal expansion (SNTE). Here we review some select cases of NTE which strictly arise from anharmonic phonon dynamics, with a focus on open perovskite lattices. We find that NTE is often present close in proximity to competing structural phases, with structural phase transition lines terminating near $T$=0 K yielding the most superlative displays of the SNTE effect. We further provide a theoretical model to make precise the proposed relationship among the signature behavior of SNTE, the proximity of these systems to structural quantum phase transitions and the effects of phase fluctuations near these unique regions of the structural phase diagram.  The effects of compositional disorder on NTE and structural phase stability in perovskites are discussed.

\end{abstract}

\maketitle

\section{Introduction}

\begin{figure}[b]
	\centering
	\includegraphics[width=\textwidth]{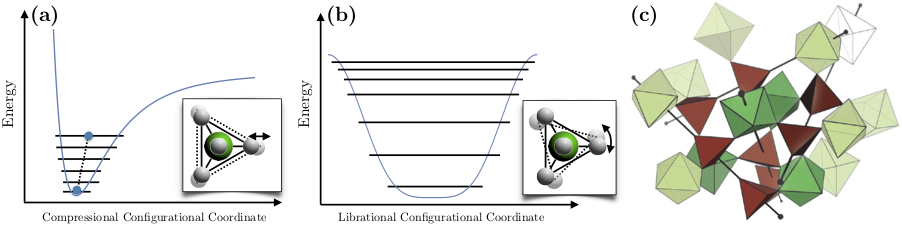}
	\caption{Intermolecular potentials provide a heuristic explanation for the common occurrence of (a) PTE and (b) NTE.  (c) The complex structure of the low-symmetry $\alpha$-phase of ZrW$_2$O$_8$, from Ref.~[\onlinecite{Hancock2004b}].}
	\label{fig:ZWO}
\end{figure}

Thermal expansion is among the most widely recognized thermodynamic properties of materials.  From a textbook perspective \cite{Ashcroft1976}, thermal expansion occurs through anharmonic free energy terms arising from nuclear lattice degrees of freedom. The dominant appearance of the positive thermal expansion (PTE) found in both research-grade and industrial materials is heuristically ascribed \cite{Barrera2005,Miller2009,Takenaka2012} to the expected anharmonic behavior of a generic interatomic potential, which is hard at short distance and soft at large distance (Figure \ref{fig:ZWO}a). As temperature is raised, higher energy excitations are populated which have an ever increasing mean separation, dilating the bond and presumably lattice dimensions. Of course this is not a theorem any more than crystals are molecules and collective motion of lattices permit various potential landscapes, such as a librational coordinate of tetrahedral molecular solids \cite{Prager1997}, which possess clear qualitative differences (Figure \ref{fig:ZWO}b).

Mention of negative thermal expansion (NTE), a material's tendency to shrink when heated, often evokes discussion of liquid water-ice expansion responsible for icebergs and the 4K temperature window above the ice-water phase boundary where phase fluctuations occur. This is an example of a route to achieving NTE which relies on broadened phase transitions between a low-temperature high-volume phase fluctuating into a high-temperature low-volume phase, other examples of which include the industrial alloy InVar \cite{Guillaume1920} (Fe$_{64}$Ni$_{36}$) and more recently discovered NTE materials \cite{Azuma2011,Takenaka2005,Qu2012a,Chen2013,Chen2013a} (for more details on this approach, see Takenaka's review in this volume \cite{Takenaka2018}). While this route to realizing NTE is promising for many applications requiring only dimensional concerns, NTE at these broadened transitions occurs only in heavily restricted regions of the magnetic and electronic phase diagrams, constraining a thermodynamic number of degrees of freedom to achieve a single mechanical characteristic.  Thus, these types of NTE materials will be severely restricted in their potential for multifunctional applications.

Remarkably, there exists a growing class of materials with strong, isotropic, robust, and thermally persistent NTE that arises from structural motifs \cite{Mary1996,Evans1997,Evans1997a,Pryde1997,Perottoni1998,Ramirez1998,Ernst1998,Ramirez2000,Mittal2001,Cao2002,Ouyang2002,Cao2003,Drymiotis2004,Mittal2004,Hancock2004,Tucker2005,Kennedy2005,Pantea2006,Keen2007,Tucker2007,Figueiredo2007,Schlesinger2008b,Keen2011,Gallington2013,Gupta2013,Gallington2014,Bridges2014,Sanson2014,Greve2010,Morelock2013b,MARTINEK1968,Zhou2008,Han2007,Chapman2005a,Mittal2018}. NTE in these systems is often discussed in connection with transverse fluctuations of a linkage between volume-defining vertices, which may accompany the librational, or hindered rotational motion of polyhedral subunits. The energy landscape for such motion tends to be much softer (0-2 THz) than bond-stretching motion (10-30 THz in oxides) which is often the implicated culprit of PTE. Here, NTE arises from the cooperative fluctuations of the bond network on THz time scales under very strong anharmonic influences and appears without necessarily constraining the magnetic or electronic phase diagram, permitting one to envisage new multifunctional materials with diverse mechanical, spin, orbital, thermal, electronic, superconducting, and more exotic order coexisting with NTE. Study of the unusual physics behind this type of NTE informs discovery efforts to find new contexts for this remarkable phenomenon. In addition, NTE materials hold promising application potential in stabilizing fiber Bragg gratings for high-speed telecommunication \cite{Kutz2002,Fleming1997}, substrates for devices which benefit from thermally controlled stresses and the formation of rigid composite structural materials with engineered thermal characteristics through combinations of PTE and NTE components \cite{Balch2004,DeBuysser2004,Lommens2005,Sullivan2005,Lind2011}.

This second circumstance for NTE, which we term \textit{structural} NTE (SNTE), is the focus of the present article.  The field of SNTE has been met with sustained interest from the physics, chemistry, and materials science communities since the re-discovery of the strong SNTE in ZrW$_2$O$_8$ in 1996 \cite{Mary1996,MARTINEK1968}. The SNTE effect here persists over the temperature range 4-1050 K and has a sizable linear coefficient of thermal expansion (CTE) of $\alpha_\ell \simeq -9$ ppm/K near room temperature, which is isotropic due to the cubic symmetry maintained at all observed temperatures under ambient pressure. The low-symmetry $\alpha$-phase structure of ZrW$_2$O$_8$ (Figure \ref{fig:ZWO}c) consists of ZrO$_6$ octahedra and WO$_4$ tetrahedra in the $P2_13$ space group, which has a screw axis along $[111]$. An order-disorder structural transition to a (cubic) $Pm\bar{3}$ $\gamma$-phase occurs at zero pressure and $T_c \simeq 450$K.  The NTE effect survives the structural transition, with a small discontinuity and reduction in the CTE to $\alpha_\ell \simeq -6$ ppm/K.  Furthermore, application of hydrostatic pressure at $T = 300$K first induces an orthorhombic transition at $P_c = 0.3$ GPa, followed by pressure-induced amorphization realized between $P = 1.5-3.5$ GPa \cite{Ravindran2001,Evans1997,Perottoni1998}.  Both the $\alpha$- and $\gamma$-phases contain four formula units, $N = 44$ atoms, in each unit cell, leading to a complex phononic structure with 3 acoustic and $3N - 3 = 129$ optical branches. 

Despite decades of intense research, the complex structure and associated dynamics of the ZrW$_2$O$_8$ lattice and the related $MA_2$O$_8$ compounds complicates the interpretation of both theoretical and experimental investigations into the mechanisms of SNTE.  For instance, a commonly identified feature in the low-temperature $\alpha$-phase is the two WO$_4$ tetrahedra with unshared ``terminal'' oxygen atoms aligned along the screw axis.  The under-constrained freedom of these tetrahedra along this axis are often cited as being responsible for the softness of the crucial NTE modes, but there is much debate as to the precise nature of the mode and its contributions to NTE \cite{Ramirez2000,Hancock2004}. Several attempts at describing the soft mode as either a translation or rotation of the WO$_4$ polyhedron were addressed via the space group symmetry - both rotational and translational motion are permitted and necessarily coupled due to the lost inversion symmetry. Another level of controversy in ZrW$_2$O$_8$ is the extent to which the molecular subunits may be regarded as rigid \cite{Cao2002,Tucker2007,Bridges2014,Dove2016,Sanson2014}. Although ZrW$_2$O$_8$ presents clear scientific challenges, its discovery is significant in that it ignited a flurry of research into the microscopic origins of the SNTE, both theoretical and experimental, employing both thermodynamic \cite{Ramirez2000} and spectroscopic \cite{Ernst1998,Hancock2004,Pantea2006,Drymiotis2004} probes of the low-energy lattice behavior.  Some essential, guiding observations were revealed during the ensuing years: (i)  ZrW$_2$O$_8$ has unusually low-energy lattice modes near 2-3meV \cite{Ramirez2000, Hancock2004}, (ii) structural phase transitions are readily induced via light hydrostatic pressure \cite{Ravindran2001,Evans1997,Perottoni1998} and (iii) the SNTE arises from a delicate balance of the degrees of freedom and constraint in the host lattice framework \cite{Cao2002,Tucker2007,Bridges2014,Dove2016,Sanson2014}.

One central question motivating SNTE research is why some materials show SNTE and others do not?  To address this question will open avenues to discovery of new NTE materials and advancing technology born from its unique properties.  While the precise mechanisms behind the dramatic SNTE in ZrW$_2$O$_8$ are still under contention, a variety of other simpler systems with equally impressive SNTE have been discovered in recent years \cite{Greve2010,Rodriguez2009,Hancock2015}.  In moving towards the goal of a deeper understanding of SNTE mechanisms, we sharpen our focus on the growing class of perovskite materials exhibiting NTE, including ScF$_3$, ReO$_3$ and related structural family members.  We consider the rich structural phase diagrams of the perovskite structure and their description in terms of octahedral tilts and the corresponding slow lattice dynamics associated with the structural transitions. Although numerous, the hierarchy of phases is well understood and documented, making perovskites a particularly simple framework on which to study the interplay of lattice dynamics and macroscopic phenomena like NTE.  In particular, we note how the corresponding dynamic modes of the perovskite lattice relate to soft-mode instabilities that accompany the approach to realized and incipient structural phase transitions and how these are coupled to mechanisms resulting in SNTE.  Most importantly, we further develop the apparent connection between the emergence of SNTE alongside phase fluctuations that occur near $T$=0K structural quantum phase transitions (SQPTs), for which we present the available experimental evidence and develop a systematic modeling scheme to explain the coupling between phase fluctuations and thermal expansion anomalies in perovskite materials.

\section{Perovskites, Structural Phases and Soft-Mode Induced Transitions}

The perovskite lattice structure may well be identified as the double-helix of the solid state - a framework which is highly functionalizable, tunable, robust, and underpins perhaps every known category of physical behavior. This includes high-temperature superconducting, itinerant ferromagnetic, local ferromagnetic, ferroelectric, insulating, metallic, glassy, as well as a plethora of antiferromagnetic and other poorly understood phases which appear to compete, coexist, and cooperate within typically rich and complex phase diagrams \cite{Kimura2002,Maekawa2004,Takagi2010,Ngai2014}.  The cubic perovskites are lattice structures with formula unit $ABX_3$, where the $A$-site is typically an alkali or alkaline earth metal ion, $B$ is a transition metal and $X$ is the anion, most commonly forming an oxide or a halide. The highest-symmetry solid phase is shown below in Figures \ref{fig:CLM}a and \ref{fig:Perov}a, with a cubic space group symmetry $Pm\bar{3}m$ and the $B$-site ions in an $n = 6$ octahedral coordination environment of $X$-site anions.  A hierarchy of structural phases in the perovskites are achieved through various concerted rotations of the $BX_6$ coordination octahedra.  These phases have been cataloged and a relationship between octahedral tilts and the lower-symmetry space groups due to these structural distortions have been developed \cite{Glazer1972, Glazer1975} and are well-known in the ferroelectric community \cite{Benedek2013}. 

\begin{figure}[h]
	\centering
	\includegraphics[width=.5\textwidth]{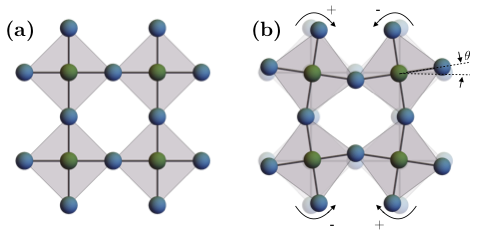}
	\caption{(a) A single 2D layer of perovskite octahedra in the high-symmetry cubic phase, viewed along one of the 3 fourfold axes and (b) the same layer with a non-zero tilt angle $\theta$, showing the constrained motion of neighboring octahedra due to shared inter-octahedral $X$-sites.}
	\label{fig:CLM}
\end{figure}

The scheme for indexing the possible octahedral tilts begins with a 2x2x2 unit cell of the cubic $Pm\bar{3}m$ perovskite and considers rotations of the octahedra about each of the 3 fourfold ($C_4$) axes of the cubic phase.  In the plane normal to a given rotation axis, neighboring octahedra are constrained to rotate at equal angles ($\theta$) of opposite sign, since neighboring $B$-sites are bonded to a common $X$-site anion (Figure \ref{fig:CLM}b); there is, however, a choice in the phase of rotations for columns of octahedra along the rotation axis.  Which phase pattern is realized is denoted by a superscript of $+$ or $-$ for in- and out-of-phase stacking, respectively, or a superscript of $0$ indicating a null rotation.  The equality of rotation angles around each axis is given by using repeated characters.  For instance, in Glazer notation $a^+b^+c^+$ represents three unequal rotations about $[100]$, $[010]$ and $[001]$, with all rotations in phase along each respective axis.  Overall, there are 23 distinct possibilities of perovskite space groups and octahedral tilting patterns, which can be cubic to triclinic and anything in between.  Several relevant examples of perovskite distortions and the Glazer notation are given in Figure \ref{fig:Perov}.

\begin{figure}[h]
	\centering
	\includegraphics[width=\textwidth]{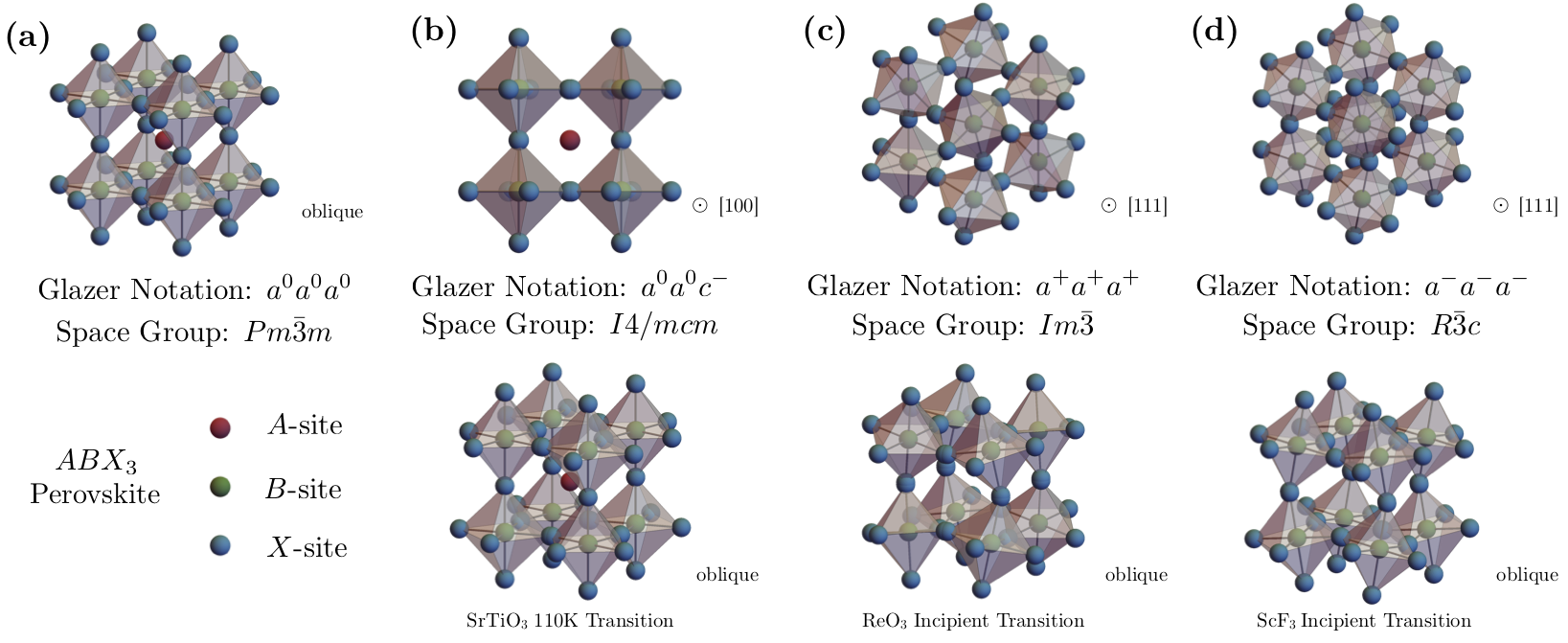}
	\caption{Shown here is (a) the general perovskite structure of formula unit $ABX_3$ (see text), in the highest symmetry cubic space group $Pm\bar{3}m$ corresponding to a Glazer tilt notation $a^0a^0a^0$.  Also shown are common octahedral tilt lower-symmetry perovskites found in (b) the SrTiO$_3$ $T_c = 110$K $Pm\bar{3}m$ to $I4/mcm$ tetragonal structural transition (c) the triply degenerate $M_3^+$ phonon condensation in the low-$T$, high-$P$ ReO$_3$ structural transition (see text) and (d) the triply degenerate $R_4^+$ phonon condensation responsible for the rhombohedral transition is $B$F$_3$ open perovskite $3d$-transition metal trifluorides which also acts as the dynamic  soft-mode rotations in ScF$_3$, corresponding to Glazer tilts of $a^0a^0c^-$, $a^+a^+a^+$ and $a^-a^-a^-$, respectively.}
	\label{fig:Perov}
\end{figure}

\begin{figure}[h]
	\centering
	\includegraphics[width=.5\textwidth]{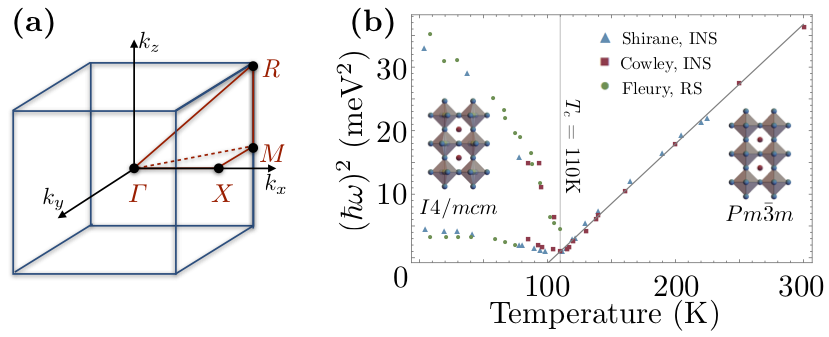}
	\caption{(a) The cubic Brillouin zone indicating the high-symmetry reciprocal lattice points $\Gamma$, $X$, $M$ and $R$.  (b) $R$-point mode softening on the approach to the $T_c \simeq 110$K structural transition in SrTiO$_3$, measured from neutron \cite{Shirane1969, Cowley1969} and Raman \cite{Fleury1968} scattering.  Black line for $T \geq T_c$ shows agreement with the predicted soft-mode frequency for a dynamically driven second-order phase transition as given in Eq. \ref{eq:Landau}.}
	\label{fig:STO}
\end{figure}

One of the best-studied structural instabilities in a perovskite structure is the  transition at $T_c \simeq110$K in SrTiO$_3$, first identified with electron spin resonance (ESR) spectra by Unoki and Sakudo \cite{Unoki1967} and later confirmed by many others \cite{Cowley1969, Cowley1964, Shirane1969, Fleury1968} via inelastic neutron scattering (INS), X-ray diffraction and Raman spectroscopy (RS).  The room-temperature structure of SrTiO$_3$ is that of the common $Pm\bar{3}m$ space group depicted in Figure \ref{fig:Perov}a, but signatures of tetragonal symmetry in the ESR and Raman \cite{Fleury1968}  spectra are observed below $T \simeq 110$K, along with anomalies in the elasticity \cite{Bell1963}.  Details of the atomic displacements reveal the lower-symmetry structure is the tetragonal $I4/mcm$ space group, which corresponds to a $[001]$-phase-staggered rotation of the TiO$_6$ octahedra about a $[001]$ rotation axis, that is an octahedral tilting pattern of $a^0a^0c^-$ (Figure \ref{fig:Perov}b). The displacements are related to the polarization of a zone-boundary optical phonon (irrep. $R_{25}$) existing at the $R$-point of cubic Brillouin zone (BZ) (Figure \ref{fig:STO}a).  In real space, the lowered-symmetry results in an effective doubling of the unit cell dimensions along one axis. In reciprocal space, however, the symmetry lowering occurs through a halving of the Brillouin zone and results in formation of new Bragg peaks as seen in an elastic scattering pattern (X-ray, neutron, electron). Dynamically, one can associate the transition to a slowing down of an optical phonon near the $R$ ($\pi\pi\pi$) point at the corner of the cubic Brillouin zone, corresponding to a ``freezing" or ``condensation" of one component of the triply degenerate $R$-point ``soft" mode.

SrTiO$_3$ is the first material in which soft modes were measured using inelastic scattering, and their concomitance with structural phase transitions was subsequently established through their observation in many other perovskites, e.g. LaAlO$_3$, KMnF$_3$, PbTiO$_3$ and BaTiO$_3$ \cite{Shirane1974}.  A \textit{soft-mode} can generally be defined as any normal mode of the dynamic lattice whose energy or, equivalently, frequency of vibration decreases anomalously.  When such a vibrational frequency reaches $\hbar \omega = 0$, the lattice becomes structurally unstable with respect to the displacements of this normal mode, and a subsequent symmetry-lowering, static deformation occurs to restore stability. For the simplest case of Landau-Ginsburg-Devonshire theory treated at the mean-field level, one expects a temperature dependence for the soft mode frequency \cite{Cowley1980, Shirane1974, Scott1974}:

\begin{equation}
	\label{eq:Landau}
	\omega_s(T) \propto \sqrt{|T - T_c|}
\end{equation}

This dependence for the $R$-point soft-mode in SrTiO$_3$ is shown in Figure \ref{fig:STO}b.  This transition can be described by an order parameter, a quantity that is zero above and develops non-zero average values below $T_c$, which follows the angle of rotation of the TiO$_6$ octahedra about the principal axis in the low-symmetry tetragonal structure.  The transition in SrTiO$_3$ is, by all experimental accounts, second-order (continuous) in nature, but for many structural phase transitions signatures of the more common first-order (discontinuous) behavior renders the soft-mode approach invalid \textit{a priori}.  Nonetheless, soft modes can be used to interpret weakly first-order transitions and their frequency can be indicative of an incipient transition due to soft-mode coupling to other, primary order parameters.  The 110K transition in SrTiO$_3$ is also a prototypical example of critical behavior that can emerge in the vicinity of a structural transition, most notably the ``central-peak" phenomenon discovered through an anomalous quasi-elastic peak in INS energy-transfer spectra, which can be explored elsewhere \cite{Halperin1976, Topler1977, Riste1993}. 

In extreme cases, a material can approach dynamic instability with lowering temperature to near-zero soft mode energy, yet no temperature-induced transition is observed.  In this situation, subsequent application of pressure, introduction of compositional disorder (doping) or other non-thermal parameters can perturb the ground-state of the system to drive the transition at $T = 0$K, realizing a quantum phase transition (QPT) \cite{Sachdev2011}.  Research surrounding the breakdown of canonical physical behavior near these quantum critical points (QCPs) is interesting in its own right \cite{Coleman2005, Gegenwart2008} but we below focus on QCPs within the structural phase diagrams and their relationship to the development of SNTE in a subset of the perovskites.

\section{NTE in Perovskite Frameworks} \label{sec:snteperovs}

\begin{figure}[b]
	\centering
	\includegraphics[width=.5 \textwidth]{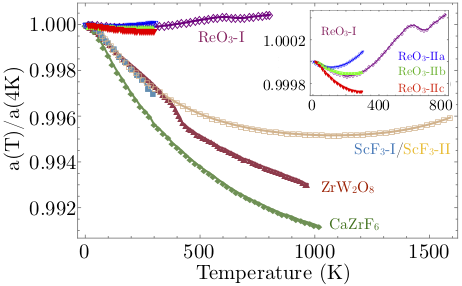}
	\caption{Shown here is the normalized cubic lattice parameter temperature dependence $a(T)/a(4K)$ for ReO$_3$, ScF$_3$ and CaZrF$_6$ NTE perovskites with strong SNTE material ZrW$_2$O$_8$ for comparison.  Inset focuses on available data for ReO$_3$ on 4 different samples, one (ReO$_3$-I) over an extended temperature range \cite{Chatterji2009b} and the other 3 (ReO$_3$-II a-c) over a smaller temperature range, investigating the effects of compositional disorder \cite{Rodriguez2009} (see Sec. \ref{sec:disorder}).  Data taken from: ZrW$_2$O$_8$ \cite{Mary1996}, CaZrF$_6$ \cite{Hancock2015}, ScF$_3$-I \cite{Handunkanda2015}, ScF$_3$-II \cite{Greve2010}, ReO$_3$-I \cite{Chatterji2009b} and ReO$_3$-II \cite{Rodriguez2009}. }
	\label{fig:LatPar}
\end{figure}

%ReO3%
Most oxide perovskites $AB$O$_3$ form with an $A$-site, otherwise requiring a rare hexavalent electronic configuration for charge balance.  One prominent exception is ReO$_3$, which forms with no $A$-site and maintains its cubic $Pm\bar{3}m$ space group symmetry down to the lowest measured temperatures.  In addition, ReO$_3$ has been known to exhibit SNTE for many years, which is often attributed to soft modes permitted by the open-perovskite ($A$-site-free) structure. The lack of the $A$-site puts fewer dynamical constraints on the motion of the ReO$_6$ octahedra in comparison to the constraints imposed by the $A$-site in other perovskites.  This permits large anisotropic thermal displacements of the linking oxygen atoms perpendicular to the Re-O-Re bond direction, making ReO$_3$ more susceptible to lattice instabilities corresponding to these octahedral tilt patterns. This openness to the structure has also been noted as a key feature in many other SNTE materials, including ZrW$_2$O$_8$, leading to a larger set of soft, low-energy phonons that have mainly been identified as the cause of SNTE. Reports on the size of the SNTE effect in ReO$_3$ vary, but in one report, SNTE was observed in two separated temperature windows of $2$ - $220$ K and $600$ - $680$ K \cite{Chatterji2009b} with a maximum measured linear thermal CTE of $\alpha_\ell = -2.56$ ppm/K \cite{Dapiaggi2009} (Figure \ref{fig:LatPar}).

\begin{figure}[t]
	\centering
	\includegraphics[scale=.65]{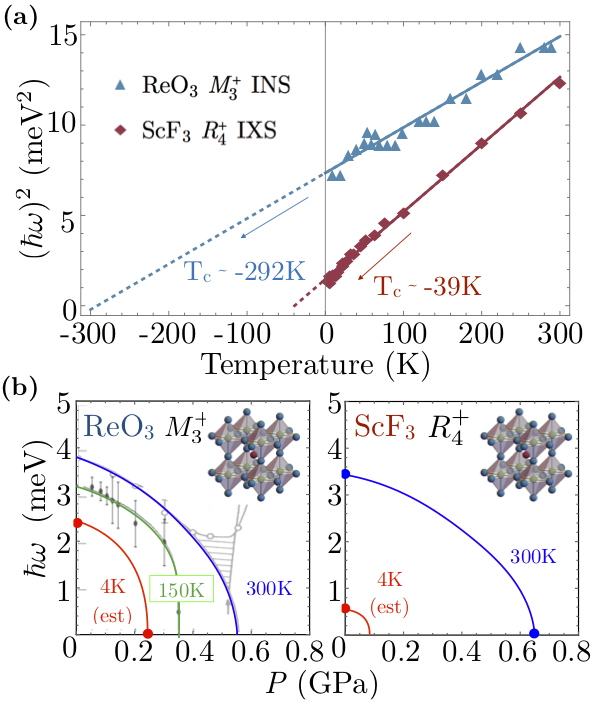}
	\caption{(a) Shows the squared energy $(\hbar \omega)^2$ of the $M_3^+$ and $R_4^+$ soft modes vs. temperature at ambient pressure in ReO$_3$ and ScF$_3$, respectively.  As indicated, extrapolation by mean-field approximation (Eq. \ref{eq:Landau}) indicates structural transition temperatures at ambient pressure of $T_c \simeq -292$K for ReO$_3$ and $T_c \simeq -39$K for ScF$_3$.  Figure (b) shows the soft mode energy versus pressure at various temperatures for (left) the $M_3^+$ mode in ReO$_3$ and (right) the $R_4^+$ mode in ScF$_3$. Solid lines are guide to the eye and symbols in both (a) and (b)  are taken from references \cite{Axe1985, Handunkanda2015, Chatterji2009}. Comparison supports our assignment that ScF$_3$ is closer to a SQPT than ReO$_3$.}
	\label{fig:NTEIncip}
\end{figure}

ReO$_3$ undergoes several structural phase transitions under hydrostatic pressure and is most studied at room temperature. Early INS investigations at ambient temperature established that ReO$_3$ undergoes a pressure-induced second-order phase transition at $P_c = 0.52$ GPa \cite{Axe1985}. Further studies of transport at $T=2$ K showed that the lowest structural phase boundary terminates at a light hydrostatic pressure of only $P_c = 0.25$ GPa, observed through a change of Fermi surface cross section \cite{Schirber1979}; however, few reports are available in this difficult $P$-$T$ region. Based on early high-temperature data, the pressure-induced phase is likely the tetragonal $P4/mbm$, although recent indications of a direct transition to a cubic $Im\bar{3}$ phase have also been reported \cite{Axe1985, Jorgensen1986}.  Neutron diffraction at elevated hydrostatic pressures revealed that the $Im\bar{3}$  phase is stable in the pressure range $0.5$ to $13.2$ GPa, above which the phase changes to the rhombohedral $R\bar{3}c$ space group \cite{Jorgensen2004}. The soft mode driving the pressure and temperature induced structural transition between the $Pm\bar{3}m$ and $Im\bar{3}$ cubic phases was shown to be three-component $M_3^+$ phonon mode involving anti-phase rotation of the neighboring ReO$_3$ octahedra in an $a^+a^+a^+$ tilt pattern (Figure \ref{fig:Perov}c).   The temperature-dependence of the $M_3^+$ mode frequency as a function of temperature at ambient pressure is shown in Figure \ref{fig:NTEIncip}b, along with a fit to the mean-field result (Eq. \ref{eq:Landau}).  This mode is significant in that it is used to understand NTE behavior of open-perovskite systems but is also identified as an order parameter of the phase transition \cite{Chatterji2009}.

%ScF3%
Unlike oxides, fluorides commonly form stable $A$-site-free perovskite structures $B$F$_3$ due to the wider array of available $B^{3+}$ ion valence configurations among the transition metals. Prominent among these open-perovskite fluorides is ScF$_3$, which was discovered in 2010 by Greve et al. \cite{Greve2010} to exhibit a robust NTE effect, which has significant maximal magnitude of the linear CTE $\alpha_\ell \sim -15$ ppm/K, persisting over the broad temperature range of $4$-$1050$K (Figure \ref{fig:LatPar}).  At room temperature, ScF$_3$ crystallizes isostructurally to ReO$_3$ with space group symmetry $Pm\bar{3}m$ and has been found to possess related structural instabilitiies corresponding to zone-boundary optical phonons.  In ReO$_3$, the condensing soft mode responsible for the low-$T$ high-$P$ structural phase transition is the $M_3^+$ distortion, while ScF$_3$ and other 3$d$-transition metal trifluorides fall into the lower-symmetry rhombohedral $R\bar{3}c$ space group symmetry, attributed to the condensation of the $R_4^+$ optical phonon.  

Although the cubic phase of ScF$_3$ is stable at ambient pressure over the entire temperature of the solid phase down to $T = 0.4$K \cite{Romao2015}, X-ray diffraction \cite{Aleksandrov2009,Greve2010} and Raman spectroscopy \cite{Aleksandrov2009} results have revealed that ScF$_3$ undergoes several pressure-induced phase transitions.  The first is from cubic to rhombohedral ($c$-$r$) after $P_c = 0.6$ GPa at $T$=300K, with a subsequent rhombohedral to orthorhombic transition occurring above $P_c = 3.0$ GPa. The $c$-$r$ transition has an observed pressure dependence of $dT_c/dP \simeq 525$ K/GPa \cite{Greve2010, Aleksandrov2009, Aleksandrov2011}. Measurement of the lattice dynamics and the soft $R_4^+$ mode responsible for the rhombohedral transition were performed using inelastic x-ray scattering (IXS), which revealed a 1D manifold of soft optical phonons that circumscribe the entire cubic Brillouin zone-edge.  At room temperature, this manifold of modes along $M$-$R$ have energy $\hbar \omega \simeq 3$ meV, softening nearly uniformly to $< 1$ meV at cryogenic temperatures (see Figure  \ref{fig:NTEIncip}b) \cite{Handunkanda2015}.  The IXS results combined with structural data permit an estimation that pressures as small as $P_c \simeq 0.074$ GPa would be sufficient to drive the transition to 0 K.  The sensitivity of the phase boundary suggests that the nature of the cubic phase is delicate at low temperature and has been shown to be susceptible to even mild perturbations \cite{Morelock2013, Morelock2014, Morelock2015}, implying that the ground state of this ionic insulator lie in close proximity to a SQPT. 

Phase stability and thermal expansion effects in the open-perovskite trifluoride structure have also been investigated thoroughly through chemical substitution.  Chemical substitutions of Sc by Ti  \cite{Morelock2014}, Al \cite{Morelock2015}, and Y \cite{Morelock2013} have been reported and the effects of this compositional disorder will be discussed in detail in Sec. \ref{sec:theory} and \ref{sec:disorder}.  Other investigations of changing the stoichiometry have resulted in a related class of hexafluoride compounds, one of which is CaZrF$_6$.  This material has $Fm\bar{3}m$ space group symmetry and is related to the $Pm\bar{3}m$ structure of ReO$_3$ but with a staggered $B$-site ion; that is, alternating CaF$_6$ and ZrF$_6$ octahedra tiling a simple cubic point-group structure. The resultant ($\pi\pi\pi$) pattern is likely a key feature when attempting to relate these materials, and is in particular likely to impact the (simple cubic) $M$-$R$  BZ edge mode dispersion and dimensional reduction observed in ScF$_3$ \cite{Handunkanda2015}. Compared to ScF$_3$, this system has isotropic NTE of larger magnitude $\alpha_\ell \simeq -18$ ppm/K over a temperature range $> 1050K$ (Figure \ref{fig:LatPar}).   At $P=0$ the system also remains cubic at all temperatures above $10$ K but a pressure-induced transition to a disordered state occurs near $P_c = 0.45$ GPa \cite{Hancock2015}, closer in the $P$-$T$ diagram than the $c$-$r$ transition in ScF$_3$ ($\sim 0.6$ GPa ) \cite{Greve2010, Aleksandrov2003, Aleksandrov2009}.  Early computational work suggests the $\Gamma-X$ manifold in the cubic BZ for this compound contribute most strongly to NTE \cite{Gupta}, but inelastic scattering measurements of the phonon dynamics are needed to assess the influence of the staggered substitution on the critical SNTE dynamics.

The open-perovskites presented above demonstrate this frameworks' favorable environment for harboring SNTE, but begs the question of why most other purely stoichiometric transition metal trifluorides and perovskites show more conventional thermal expansion.  The lattice parameters of each of these SNTE perovskites are plotted in comparison to the prototypical SNTE material ZrW$_2$O$_8$ in Figure \ref{fig:LatPar}, which gives a clear ranking of SNTE perovskites by the magnitude of the NTE effect ((1) CaZrF$_6$, (2) ScF$_3$, (3) ReO$_3$).  This ranking is the opposite ordering one gets in terms of pressure required to induce the structural phase transition nearest ambient conditions, a correlation suggestive that proximity to a SQPT and strength of NTE are interrelated.  We demonstrate this point for ReO$_3$ and ScF$_3$ in Figure \ref{fig:NTEIncip}.  These plots consider the soft-mode in each system, the $M_3^+$ phonon in ReO$_3$ and the $R_4^+$ in ScF$_3$ and the available data for the energy of these modes as a function of pressure and temperature. First considering the $T$-dependent data at $P = 0$ in Figure \ref{fig:NTEIncip}a, extrapolation of the squared mode energies by Eq. \ref{eq:Landau} provides a quantitative measure of the proximity to a dynamically-driven SPT, yielding $T_c \simeq -292$K in ReO$_3$ \cite{Chatterji2009} and $T_c \simeq -39$K in ScF$_3$ \cite{Handunkanda2015}. Furthermore, isothermal measurements of the soft-mode energy versus hydrostatic pressure are provided in Figure \ref{fig:NTEIncip}b, showing that decreasing temperature and increasing pressure in both systems trend towards a QCP.  

Together, the results of these data clearly show in all respects that ScF$_3$ is closer to a SQPT than ReO$_3$.  Although data at this level in unavailable for CaZrF$_6$, the amorphization boundary at $P_c \leq 0.45$ GPa and $T = 300$K is a lower pressure threshold for the pressure-induced transitions at $300$K in both ReO$_3$ ($\sim 0.55$ GPa) and ScF$_3$ ($\sim 0.65$ GPa).  It is thus likely that ground-state of this compound is the closest to a structural instability at cryogenic temperature, while also exhibiting the most superlative SNTE effect in this class.   Our central hypothesis in the context of the materials described is that the $T=0$K termination of a structural phase boundary defines a structural quantum critical point (SQCP) where strong geometrical fluctuations associated with octahedral tilts drives NTE. In our view, the significance of the SQCP is a flattening of the energy landscape with respect to transverse fluctuation of the linkage unit: O in ReO$_3$, and F in ScF$_3$ and CaZrF$_6$.  It is worth noting that NTE arising from phase fluctuations and the displacements of a low-$T$ soft-mode is not unique to the antiferrodistortive (zone-boundary) phonons in perovskites, but has also predicted SNTE in materials with broadly distinct structures and geometrical motifs, e.g. the Hg dimer in Hg$_2$I$_2$ \cite{Occhialini2017hg2i2}, the CN molecule in Prussian blue analogs and related compounds\cite{Goodwin2008,Fairbank2012,Mittal2009}.  

NTE is often understood through the response of the phonon spectrum to the application of hydrostatic pressure, which has been formalized in the quasi-harmonic approximation (QHA) known as the Gr\"{u}neisen approach \cite{Ashcroft1976}.  Each phonon in the Brillouin zone of frequency and wavevector $(\omega_i, \bm{k}_i)$ is assigned a mode Gr\"{u}neisen parameter $\gamma_i$, defined as,
\begin{equation}
\gamma_i \equiv -\frac{\partial \ln \omega_i}{\partial \ln V} \equiv  \frac{1}{\kappa} \frac{\partial \ln \omega_i}{\partial P}
\end{equation}
where $\kappa$ is the isothermal compressibility.  Performing an average over all $\mathbf{k}$, weighted by the mode contribution to the heat capacity $c_{V,i}$, gives the overall lattice Gr\"{u}neisen constant $\gamma$ which is thermodynamically proportional to the volumetric thermal expansion $\alpha_V$ for isotropic materials.  At low-temperatures, the thermodynamic properties are dominated by contributions from the lowest energy excitations.  If the low-energy phonon spectrum has large magnitude, negative mode Gr\"{u}neisen parameters (negative contributions to CTE), then the $\mathbf{k}$-averaged CTE will decrease as temperature is lowered.  If strong enough to overcome the many high-energy excitations commonly attributed to conventional PTE, the overall expansion may turn negative in sign, strengthening at lower temperature, which is the typical functional form among the SNTE perovskites (Figure \ref{fig:LatPar}), before relaxing and limiting to a thermodynamically-required $\alpha_V = 0$ as $T \to 0$K.  From this viewpoint, soft-modes with NTE contributions are natural candidates for inducing overall NTE, since their energy softens with lowering temperature, enhancing the mode occupation and weighted contributions to the thermodynamics at low-$T$ in comparison to thermally-stable low energy excitations. 

In the SNTE perovskites and other SNTE materials like ZrW$_2$O$_8$, these lowest energy lattice excitations are commonly attributed to quasi-rigid dynamics of polyhedral subunits \cite{Dove2016}, i.e. the geometrically rigid octahedra as shown in Figure \ref{fig:CLM} which could correspond to $BX_6$ octahedra in ScF$_3$, ReO$_3$ or CaZrF$_6$.  These rigid unit mode (RUM) analyses model rigidity by freezing out portions of the phonon spectrum, such as high-energy bond-stretch and internal polyhedral bond-bend modes that are commonly attributed to causing PTE.  For ScF$_3$ and ReO$_3$, the antiferrodistortive, zone-edge soft modes have an interpretation as RUMs.  Moving beyond the commonly employed QHA and Landau mean-field approaches, we make the hypothesized relationship among soft RUMs, phase fluctuations and the development of SNTE precise within a systematic model in Sec. \ref{sec:theory} below.

\section{Theory of SNTE from RUM Fluctuations} \label{sec:theory}

The purpose of this section is to present a microscopic description of NTE arising from soft modes in ReO$_3$-type lattice structures. Such modes break the  symmetry of the lattice and lead to displacive structural phase transitions \cite{Giddy1993a}.  Typical examples are the $R_4^+$ mode at the  point $(1, 1, 1)\left( \pi/a \right)$ of the Brillouin zone of the cubic (c) $Pm3m$ phase in MF$_3$ (M=Sc, Al, Cr, V, Fe, Ti)  metal fluorides which upon condensation gives rise to a rhombohedral (r) $R\overline{3}c$ lattice structure and the $M_3^+$ mode at $\left(1, 1, 0 \right) \left( \pi/a \right) $ in ReO$_3$ which generates a tetragonal ($P4/mbm$) phase.

The structural transitions observed in these materials are generally described by Landau theories \cite{Axe1985, Corrales2017a}. Typically, they include an order parameter (OP) associated with cooperative tilts of a rigid unit (e.g. the MF$_6$ octahedron in the metal fluorides) coupled to long-wavelength acoustic phonons that generate volume, deviatoric and shear strains. While such mean field theories provide a fair description of the structural transitions, they fail to describe NTE, e.g.,  they predict zero thermal expansion in their high-$T$ cubic $Pm3m$ phase.

Here, we present a microscopic phenomenology that describes  NTE in these open perovskite frameworks.  The model includes the usual rigid tilts coupled to long-wavelength strain-generating acoustic modes as well as a cooperative interaction between tilts that drives the structural transition, e.g., dipolar interactions in the metal trifluorides \cite{Chen2004a, Chaudhuri2004a, Allen2006a}.  Our main result is that any solution of the model must include fluctuations of the OP to generate NTE.  We illustrate this within a so-called self-consistent phonon approximation (SCPA) in which single site fluctuations are considered while inter-site fluctuations are neglected.  This point has been appreciated before \cite{Volker82a}, however, no systematic approach has been constructed so far. 
In addition, our model allow us to parametrize measured macroscopic quantities in terms of microscopic parameters, which provides guidance for materials design.  Our model closely follows those of the well-known antiferrodistortive transitions of SrTiO$_3$ and LaAlO$_3$ \cite{Feder1970a}, with the important distinction that we include hydrostatic pressure and account for compositional disorder. The latter is aimed at describing compounds with tunable NTE through composition such as mixed solid solutions of metal triflurides \cite{Morelock2014, Morelock2015, Morelock2013b}.  For concreteness, we will consider a c-r transition similar to that in Sc$_x$Ti$_{1-x}$F$_3$ in which the threefold zone-boundary $R_4^+$ phonon splits into a low-energy $E_g$ doublet and a high energy $A_{1g}$ singlet at a transition temperature $T_c$ \cite{Daniel1990b}.

Our model analysis is by no means exhaustive. More elaborate descriptions that go beyond the picture of rigid tilts involving, for instance, distortions and translations of such building units are usually needed to describe the observed TE \cite{Li2011a}.  Also the observed structural transitions are frequently of first-order, which we do not consider here for the sake of simplicity. Nonetheless, our semi-analytic approach accounts for microscopic aspects of the phonon dynamics and its relation to NTE and help in finding general trends of the solution. Moreover, it provides the basis to build other frameworks that capture atomistic details such as first-principles-based effective model Hamiltonians \cite{ModernPerspective2007a}.

\subsection{Model Hamiltonian}

We consider a cubic lattice with  $N$ sites and choose normal mode coordinates  ${\bm Q}_i=(Q_{ix}, Q_{iy}, Q_{iz})$ in the unit cell $i\,(i=1,2,....,N)$ associated with the $R_4^+$ mode, the condensation of which leads to the  $R\bar{3}c$ rhombohedral phase.  ${\bm Q}_i$ is proportional to the local displacements generated by the cooperative tilts of the MF$_6$ octahedra. In addition, we introduce the strain tensor in Voigt notation $\epsilon_{i \alpha}$, $\alpha = 1,...,6$ in unit cell $i$, which is induced by displacements ${\bm u}_i = (u_{ix}, u_{iy}, u_{iz})$ of the centers of mass of the unit cells with respect to the acoustic-branch phonons.  In order to determine the optical phonon contribution to the thermal expansion, we must couple the displacements ${\bm Q}_i$ with strains $\epsilon_{i\alpha}$, leading to a 3-term Hamiltonian of the form,

\begin{align}
	\label{eq:Hamiltonian}
	H = H_Q + H_\epsilon + H_{Q \epsilon}.
\end{align}

Here, $H_Q$ accounts for harmonic and anharmonic energy contributions from the soft optical phonon up to quartic order in ${\bm Q}_i$; $H_\epsilon$ is the strain-induced energy depending on the elasticity through the bulk modulus $C_a$, shear moduli and hydrostatic pressure $P$; and $H_{Q \epsilon}$ models the coupling between these displacements and strain degrees of freedom with $g_a$ the coupling constant between the displacements and the volume strain. The explicit form of these terms is given in the supplementary material (SM).  To solve the statistical mechanical problem posed by the Hamiltonian in Eq. \ref{eq:Hamiltonian}, we use a variational formulation of a SCPA, in which the temperature and pressure dependence of the phonon energies $\Omega_{\nu},~(\nu=R_4^+, A_{1g}, E_g)$, displacements, strain order parameters and phase fluctuations are determined self-consistently from the minimization of the free energy \cite{Pytte1972a}.  We here focus on the main results.  The details of the model Hamiltonian and its approximate solution are given in the SM. 

\subsection{Thermal expansion, CTE, and Gr\"{u}neisen parameters.}
\begin{figure}[htp]
	\centering
		\includegraphics[scale=0.6]{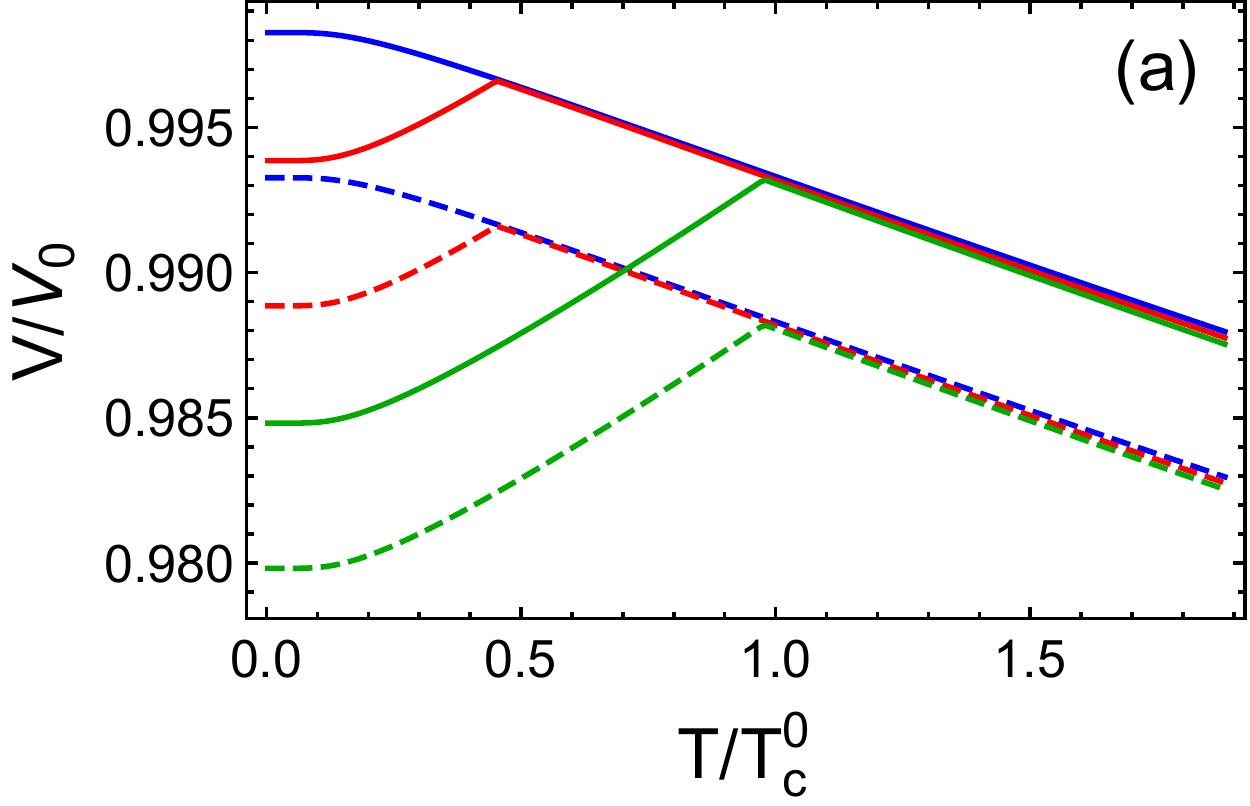}
		\includegraphics[scale=0.6]{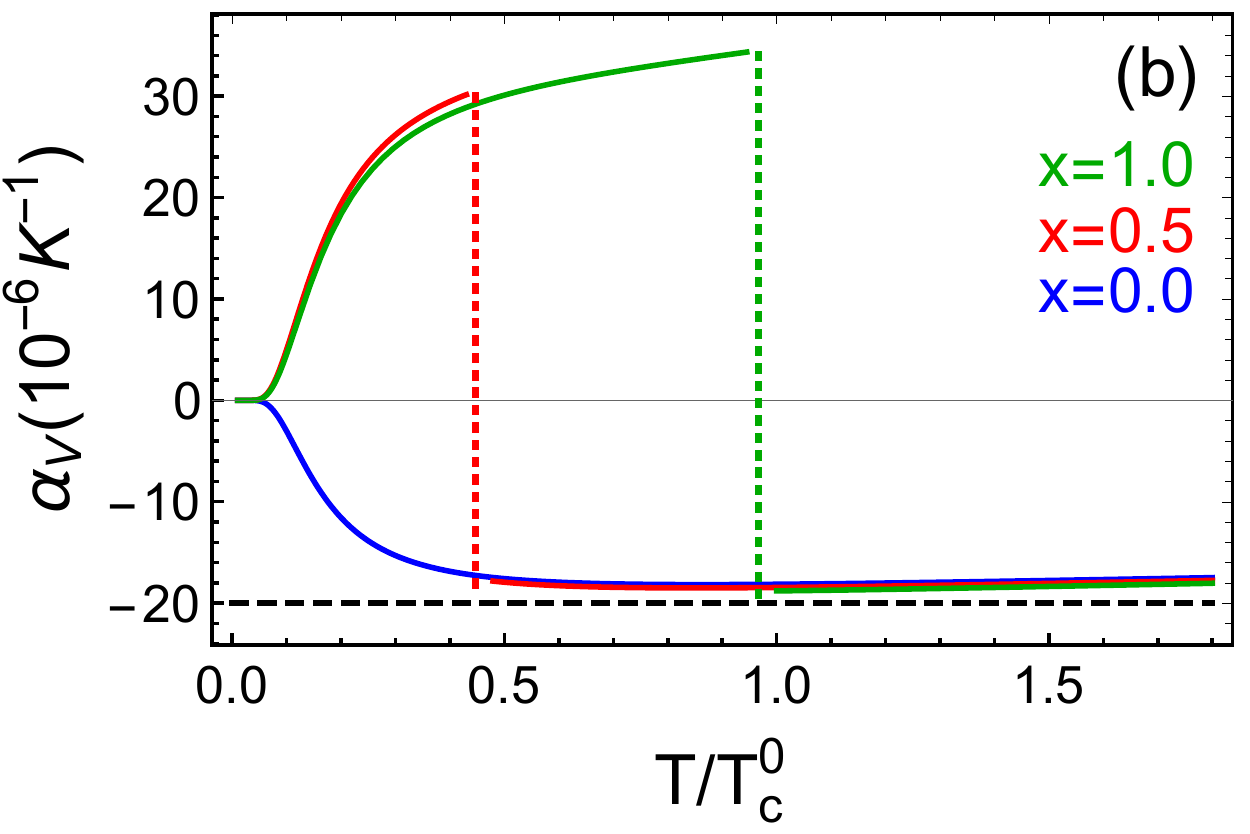}
		\includegraphics[scale=0.6]{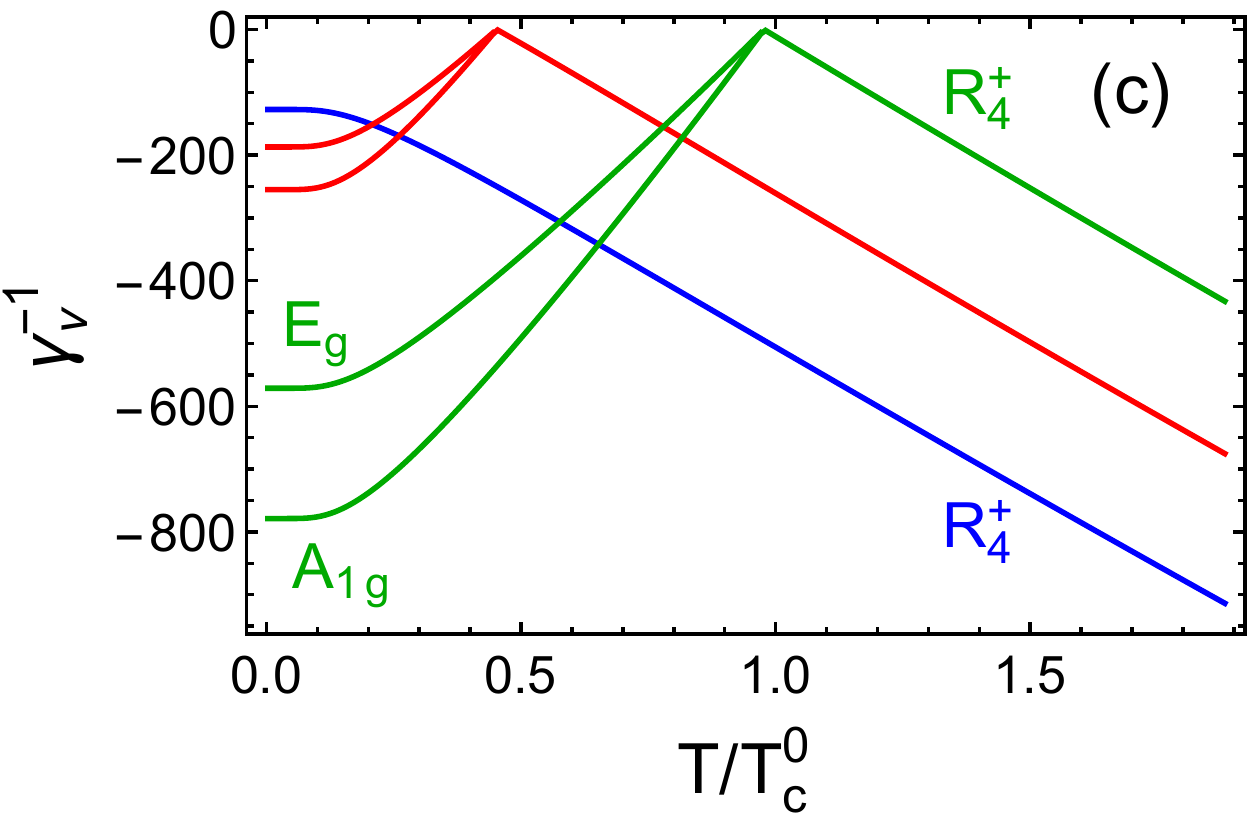}
	\caption{Calculated temperature dependence of (a) volume change, (b) CTE, and (c)
Gr\"{u}neisen parameters for $P/P_0=0$ (solid lines) and $9.0 \times 10^{-5}$ (dashed lines) for Sc$_{1-x}$Ti$_x$F$_3$. $P_0$ is the pressure needed to induce the c-r transition in ScF$_3$ ($x=0$) at $0\,$K; $T_c^0=340\,$K is the transition temperature of TiF$_3$~($x=1$) at ambient pressure. Black dashed line in (b) is the predicted CTE in the classical limit according to Eq.~(5). The phonon symmetry labels for $x=0.5$ in (c) are the same as those for $x=1$ and are not shown for clarity.}
	\label{fig:nte}
\end{figure}

We first focus on the volume strain $\bra  \epsilon_V \ket = \bra \epsilon_1 +\epsilon_2+\epsilon_3  \ket $, which gives the change in volume with temperature and pressure with respect to a reference volume $V_0$.  We use the notation $\bra .... \ket$  to denote thermal average.  By minimizing the free energy associated with the Hamiltonian in Eq. \ref{eq:Hamiltonian}, we find that the volume strain is given as follows,

\begin{align}
	\label{eq:nte}
	\bra \epsilon_V \ket = \frac{\Delta V}{V_0} =  - \frac{g_a}{C_a}  \bra \left| {\bm Q} \right|^2 \ket   - \frac{P}{C_a},
\end{align}

where $\bra \left| {\bm Q} \right|^2 \ket $ is the thermal average of the squared magnitude of the MF$_6$ tilt.  Eq. \ref{eq:nte} already illustrates one of the the main points of our work: in a mean-field theory and in the absence of pressure, $\bra \left| {\bm Q} \right|^2 \ket=0$ above $T_c$; thus fluctuations around the OP must be included to describe NTE. For instance, within the SCPA and for temperatures much greater than the phonon energy, we find that $\bra \left| {\bm Q} \right|^2 \ket \propto T$ in the cubic phase and Eq. \ref{eq:nte} gives,

 \begin{align}
	\label{eq:alphaV}
 		 \frac{\Delta V}{V_0} \simeq  \alpha_V T - \frac{P}{C_a}, ~~~~\alpha_V = - \frac{3 g_a k_B}{C_a  v_R}, 
\end{align}

where $\alpha_V$ is the CTE at high temperatures and $v_R$ is the strength of the cooperative interaction.  Figures \ref{fig:nte}~(a) and (b) show, respectively, our results for the volume change obtained from Eq. \ref{eq:nte} and its CTE ( $\alpha_V = d \bra \epsilon_V \ket / dT $) in the full temperature range. Model parameters were obtained by fitting to experiments \cite{Handunkanda2015, Morelock2014} and are given in the SM.  Despite its simplicity, 
our model produces the observed trends ~\cite{Morelock2014, Handunkanda2015}: NTE with a nearly linear $T$ dependence in the c-phase, except near $0\,$K; PTE in the r-phase; and a discontinuity in $\alpha_V$ at the phase transition. Quantitatively, the model is in good agreement in the c-phase, but $\alpha_V$ is about and order 
of magnitude less than the observed one in the r-phase.  We attribute this to having neglected the first-order character of the transition and additional phonons along the M-R line of the BZ which are known to contribute to the NTE ~\cite{Roekeghem2016a}.

We note that Eq. \ref{eq:alphaV} gives $\alpha_V$ in terms of the microscopic model parameters. It shows that mechanically compliant materials with low bulk moduli ($C_a$) and strong strain-phonon couplings ($g_a$) favor thermal expansion.  $\alpha_V$ also increases by weakening the strength of the cooperative interaction $v_R$ at the expense of decreasing the transition temperatures since $T_c \propto v_R$, as it is shown in the SM.  It also shows that the sign of this coupling plays an essential role in the thermal expansion: $g_a>0$ for NTE while $g_a<0$ for PTE. 

Another physically relevant quantity is the the Gr\"{u}neisen parameter $\gamma_{\nu}$ associated with each lattice mode $\nu=R_4^+, A_{1g}, E_g$.  We find that the temperature and pressure dependence of  $\gamma_{\nu}$ is entirely determined by the phonon energy $\Omega_{\nu}$,

\begin{align}
	\label{eq:grun}
	\gamma_{\nu}  = - \frac{g_a}{\Omega_{\nu}^2},
\end{align}

and thus diverges near the c-r transition as $\Omega_{\nu} \to 0$.  This is in agreement with previous analytic work \cite{Volker82a} and ab-initio calculations, where large, negative values for $\gamma_{R_4^+}$ have been found for ScF$_3$ ~\cite{Roekeghem2016a, Liu2015a, Li2011a}.  Figure \ref{fig:nte}~(b) shows that $\gamma_{\nu}^{-1} \propto - \left| T-T_c \right|$ at the onset of the phase transition for $x=0.5, 1.0$ and thus matches the result from Landau theory. For $x=0$, there is no transition and the deviations from linear behavior are due to zero-point fluctuations.

\section{The Role of Disorder in Perovskite SNTE materials} \label{sec:disorder}

Disorder is an inevitable part of any real material system. Here we discuss and develop the role of disorder in on the SNTE effect within the open perovskite structural class.

ReO$_3$ has been known as a SNTE material for many years, but there are varying reports of the strength and also extent in temperature over which the effect occurs, which is summarized for recent data by Chatterji \cite{Chatterji2006, Chatterji2009b} and Rodriguez \cite{Rodriguez2009} in Figure \ref{fig:LatPar} inset. Generally, ``open" perovskite oxides are rare due to the requirement of a hexavalent $B$-site and controlled substitutional studies have not been reported to our knowledge. However, the controlled disorder study by Rodriguez \cite{Rodriguez2009} compared crystals synthesized using different growth techniques and clearly showed that the highest quality crystals grown by chemical vapor transport method exhibited the largest and most thermally persistent SNTE effect. As with the physical properties of many perovskite oxides, controlled post-growth annealing procedure studies may be need to be developed to ensure the optimal NTE effect even in studies of its fundamental causes.

\begin{figure}[h]
	\centering
			\includegraphics[scale=0.9]{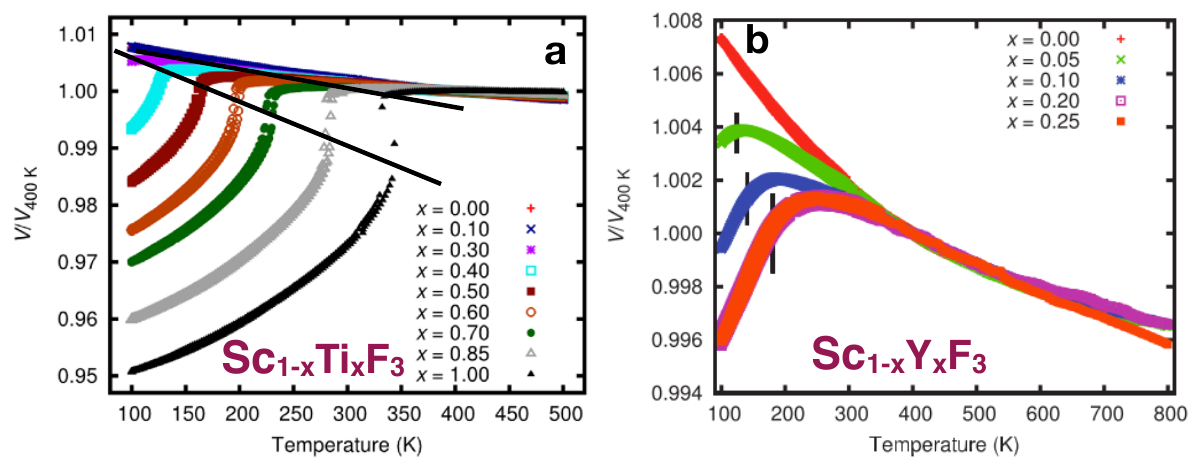}
	\caption{Compositional dependence of the SNTE effect in Sc$_{1-x}$($L = $ Ti,Y)$_x$F$_3$ for $L =$ Ti \cite{Morelock2014} and $L = $Y \cite{Morelock2013b}.  (a) Reprinted with permission from C. R. Morelock et al., Chem. Mater. 26, 1936, (2014). Copyright (2018) American Chemical Society.  (b) Reprinted from C. R. Morelock et al. J. Appl. Phys. 114, 213501 (2013), with permission of AIP publishing.}
	\label{fig:ctedoping}
\end{figure}

\begin{figure}[h]
	\centering
			\includegraphics[scale=0.7]{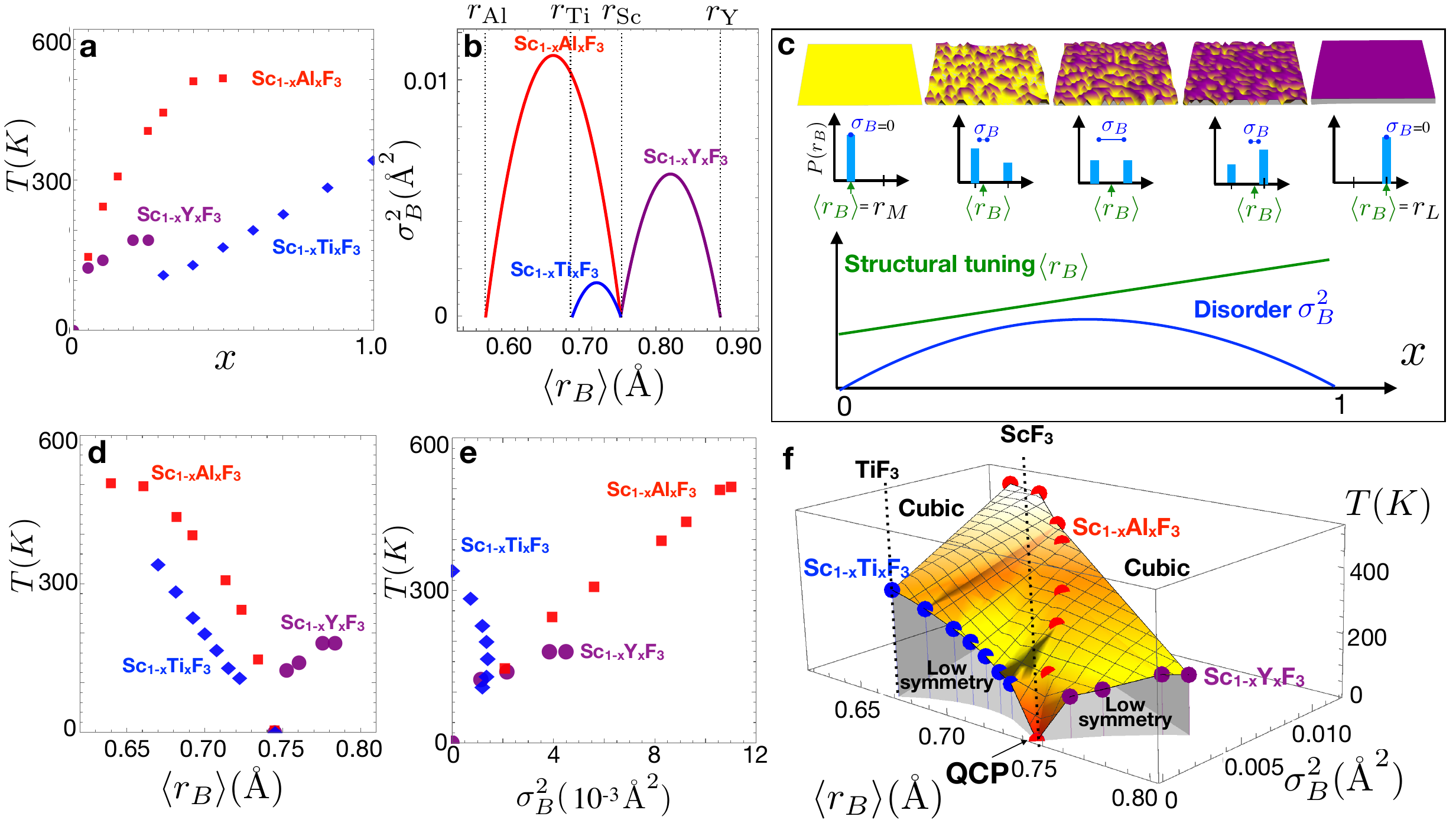}
	\caption{(a) Cubic-to-rhombohedral transition temperatures $T_c$(x) in a isovalent substitutional series Sc$_{1-x}L_x$F$_3$. (b) The dependence of the mean and variance using equation \ref{eqn:rsig} and the Shannon ionic radii of the Ti, Y, and Al. (c, top row) Illustration of the effect of substitution on the potential energy landscape in a solid solution as the composition $x$ is varied. (c, middle row) The binary probability distributions $P(r_B)$ for the five composition levels with the mean and standard deviation $\sigma_B$ of the distributions indicated by arrows and dumbbells, respectively. (c, bottom) plot of the composition dependence of the first two moments of the distribution according to equations \ref{eqn:rsig}. Plot showing the dependence of the transition temperatures as a function of (d) $\langle r_b\rangle$ and (e) $\sigma_B^2$, respectively, which show no obvious universal trend. (f) Combined plot of the transition temperatures versus the structural tuning parameter $\langle r_b\rangle$ and the disorder parameter $\sigma_B^2$.}
	\label{fig:disorder}
\end{figure}

ScF$_3$ is an unusually clean material - single crystals have been synthesized with 0.002 degree mosaic \cite{Handunkanda2016}, free of color centers, with high chemical and isotopic purity with readily available components. The flexibility afforded by the trivalent $B$-site in the trifluorides permits wide chemical tunability and provides new opportunities to observe disorder effects on SNTE. So far, the most thorough and complete studies of the substitutional series Sc$_{1-x}L_x$F$_3$ have been performed with high inorganic synthesis and high quality structural synchrotron and neutron scattering efforts of the Wilkinson group at Georgia Tech. In a series of papers \cite{Morelock2013b,Morelock2014,Morelock2015}, substitutions of $L$=Al,Y,Ti have been reported, particularly the behavior of the cubic-to-rhombohedral phase boundary in this system upon these isovalent substitutions (Figures \ref{fig:ctedoping} and \ref{fig:disorder}a). Here we develop a combined analysis of these data which permits conclusions regarding the interaction of disorder and the SNTE effect.

Following the spirit of Attfield, who has studied compositional disorder effects on the $A$-site of transition metal oxide phase transitions \cite{Attfield1998,Attfield2001,Attfield2002}, we borrow the hypothesis that the ionic radius of the substituted ions represents a local energetic influence on the stability of the ordered phase and discuss in our case the probability distribution $P(r_B)$ of finding a $B$-site ion of radius $r_B$ in the series Sc$_{1-x}L_x$F$_3$. We calculate the first two moments of this distribution and associate the mean ionic radius (1st moment) $\langle r_B\rangle$ to an energetic effect on the transition and the variance (2nd moment) $\sigma_B^2$ of the distribution as representative of disorder. For the simple binary distributions shown in Figure \ref{fig:disorder}c, these quantities are simply calculated from the nominal composition $M_{1-x}L_x$F$_3$:

\begin{eqnarray}
\label{ }
\langle r_B\rangle &=&r_M (1-x) + r_L x\\
\sigma_B^2 &=&x(1-x)(r_L-r_M)^2 \label{eqn:rsig}\\
&=&(r_M-\langle r_B\rangle)(\langle r_B\rangle-r_L).
\end{eqnarray}

These relations are general for any binary mixture, and are applied for $M$=Sc and $L$=Y,Al,Ti in Figure \ref{fig:disorder}b using the Shannon ionic radius for these trivalent ions. Appropriately, $\sigma_B^2$ is zero for the endpoints of the compositional series and is maximum at the 50-50 composition as expected in all cases. Note that for $L$=Ti, the ion best size matched to Sc, this maximum is small and the effects of disorder are expected to be weaker than for other substitutions, whereas for the much larger Y and much smaller Al ions, disorder increases substantially throughout these series. Further, substitutions of Y have opposite effects on $\langle r_B \rangle$ than substitutions of Ti and Al, therefore the three substitutional series cover well the transition in terms of both energetics and disorder.

Figures \ref{fig:disorder}d,e show the transition temperatures plotted as a function of $\langle r_B\rangle$ and $\sigma_B^2$. There is not a clear common trend in either plot, except that the Y and Al substitution series are linear in $\sigma_B^2$, implying that quenched disorder is the dominant contribution toward driving the transition, as we have pointed out previously \cite{Handunkanda2015}. For the substitution $L$=Ti, the transition temperature is linear in $x$, suggesting a dominantly energetic effect, as hypothesized based on its similar size and treated theoretically in the weak-disorder limit of the last section. Figure \ref{fig:disorder}f shows a combined plot of all three series $L$=Ti,Al,Y as a function of the structural tuning parameter $\langle r_B\rangle$ and the disorder parameter $\sigma_B^2$. This generalized disorder-energy analysis unifies the compositional dependencies of three different series with important implications, showing that disorder is deleterious to SNTE and that ScF$_3$ is situated in a very special place which is difficult to reach in the presence of any disorder. These conclusions and the known variation in the SNTE effect of ReO$_3$ indicate that disorder generally suppresses the SNTE effect and that careful work optimizing this property with respect to sample history may be necessary in some cases.

ScF$_3$ has the most dramatic SNTE effect of all members of these series and also appears at a QCP in the diagram of Figure \ref{fig:disorder}f. Figure \ref{fig:ctedoping} reproduces the figure panels for thermal expansion in each series and shows that strong SNTE persists above the transition for light substitutional levels, but weakens in all cases. We point out that no known materials exist in the large $\langle r_B\rangle$, small $\sigma_B^2$ limit, but if such a composition could be produced, would be of high interest toward exploring the robustness of SNTE to disorder. Furthermore, routine structural refinement experiments performed at liquid helium temperatures would help immensely toward refining the QCP in these systems where SNTE seems to arise near the $T$=0 termination of a structural phase boundary.

\section{Summary}

We have discussed the broad issue of SNTE with particular focus on perovskite-structured SNTE materials. We have identified the presence of several competing octahedral tilt instabilities occurring near the zero-temperature state of these materials and their associated fluctuations in the high-symmetry phase as key to the SNTE effect in these materials. We have provided a model treatment beyond mean field theory to account for these fluctuations and identified key elements that move toward control of negative thermal expansion and may be invoked for rational design and discovery of future SNTE systems. We also find that quantum mechanical effects are non-negligible and play an important role in SNTE. Finally, we have described existing data in a new analysis which attempts to isolate the influences of energetics and disorder and presented a holistic and generalizable approach leading to the conclusion that disorder disrupts the balance which drives the SNTE effect in ScF$_3$, ReO$_3$, and other SNTE materials.  Our thorough combined analysis of the physical properties and special circumstances in this simple structural class has identified trends and influences that we hope will guide discovery of new SNTE materials.

\section*{Conflict of Interest Statement}
The authors declare that the research was conducted in the absence of any commercial or financial relationships that could be construed as a potential conflict of interest.

\section*{Author Contributions}
CAO, SUH and JNH wrote sections 1-3 and 5. GGGV developed the modeling in section 4 and the SM.  All authors contributed to writing and revising the manuscript and figures.  

\section*{Funding}
Work at the University of Connecticut was provided by National Science Foundation Award No. DMR-1506825 with additional support from the US Department of Energy, Office of Science, Office of Basic Energy Sciences, under Award No. DE-SC0016481. Work at the University of Costa Rica is supported by the Vice-rectory for Research
under project no. 816-B7-601, and work at Argonne National Laboratory is supported by the U.S. Department of Energy, 
Office of Basic Energy Sciences,  Material Sciences and Engineering Division under contract no. DE-AC02-06CH11357.

\section*{Acknowledgments}
The authors would like to acknowledge valuable conversations with Peter Littlewood, Richard Brierley, Premala Chandra, and Alexander Balatsky.  GGGV acknowledges Churchill College, the Department of Materials Science and Metallurgy and the Cavendish Laboratory at the University of Cambridge where part of this work was done.

\newpage
\pagebreak
\widetext
\begin{center}
\textbf{\large Supplmentary Material: Negative thermal expansion in open perovskites near the precipice of structural stability}
\end{center}
%%%%%%%%%% Merge with supplemental materials %%%%%%%%%%
%%%%%%%%%% Prefix a "S" to all equations, figures, tables and reset the counter %%%%%%%%%%
\setcounter{equation}{0}
\setcounter{figure}{0}
\setcounter{table}{0}
\setcounter{page}{1}
\setcounter{section}{0}
\makeatletter
\renewcommand{\theequation}{S\arabic{equation}}
\renewcommand{\thefigure}{S\arabic{figure}}
\renewcommand{\bibnumfmt}[1]{[R#1]}
\renewcommand{\citenumfont}[1]{R#1}
%%%%%%%%%% Prefix a "S" to all equations, figures, tables and reset the counter %%%%%%%%%%

\maketitle

\section{Model Hamiltonian}

We consider a cubic lattice with  $N$ sites and 
choose normal mode coordinates that describe local 
displacements ${\bm Q}_i=(Q_{ix}, Q_{iy}, Q_{iz})$ 
in the unit cell $i\,(i=1,2,....,N)$ that are associated with the relevant soft $R_4^+$ phonon mode, 
the condensation of which leads to a structural transition to a rhombohedral $R\overline{3}c$ phase.
In addition, we introduce symmetry adapted strains $\epsilon_{i a} =  \epsilon_{i 1} + \epsilon_{i 2} + \epsilon_{i 3}, \epsilon_{i t} = \left( 2 \epsilon_{i 3} - \epsilon_{i 2} - \epsilon_{i 1} \right) / \sqrt{3} $ and $\epsilon_{i o} = \epsilon_{i 1} - \epsilon_{i 2}$, as well as 
shear strain components $\epsilon_{i 4}, \epsilon_{i 5}$, and $\epsilon_{i 6}$ in the usual Voigt notation: 
$ \epsilon_{i \alpha}  =  \partial u_{i \alpha} / \partial x_\alpha~(\alpha=1,2,3)$,  $ \epsilon_{i 4} = 2 \left( \partial u_{i y} / \partial z +  \partial u_{iz} / \partial y \right),   \epsilon_{i 5} = 2 \left( \partial u_{i x} / \partial z +  \partial u_{i z} / \partial x \right)$, and  $ \epsilon_{i 6} =  2 \left( \partial u_{i x} / \partial y +  \partial u_{i y} / \partial x \right)$.
${\bm u}_i = \left( u_{ix}, u_{iy}, u_{iz} \right) $ is the displacement of the center of mass of the unit cell $i$ from its equilibrium position due to the  acoustic phonon mode.
Physically, $\epsilon_{i a}$ and $\epsilon_{i t/o}$ are, respectively, volume and tetragonal strains.
We consider the model Hamiltonian,

\begin{align}
	\label{eq:SHamiltonian}
	H = H_Q + H_\epsilon + H_{Q \epsilon},
\end{align}
where,
\begin{align}
 H_Q  =  \frac{1}{2} \sum_{i, \lambda }  \Pi_{i \nu }^2  +  \frac{1}{2}  \sum_{i, \lambda}  \kappa_i Q_{i \nu}^2 + \frac{\gamma_1}{4}  \sum_{i, \lambda \lambda^\prime} Q_{i \lambda}^2  Q_{i \lambda^\prime}^2 
			+\frac{\gamma_2}{2} \sum_{i, \lambda \neq \lambda^\prime}  Q_{i \nu}^2 Q_{i \nu^\prime}^2   
			- \frac{1}{2} \sum_{ij,\lambda \lambda^\prime} Q_{i \lambda} v_{ij}^{ \lambda \lambda^\prime } Q_{j \lambda^\prime}, 
 %H_Q  =  \frac{1}{2} \sum_i \left|  {\bm \Pi }_i \right|^2 +  \frac{\kappa}{2}  \sum_{i}   \left| { \bm Q }_i \right|^2  + \frac{\gamma_1}{4}  \sum_i \left| { \bm Q }_i \right|^4 
%			+\frac{\gamma_2}{4} \sum_{i, \lambda \neq \lambda^\prime}  Q_{i \lambda}^2 Q_{i \lambda^\prime}^2   
%			- \frac{1}{2} \sum_{ij,\lambda \lambda^\prime} Q_{i \lambda} v_{ij}^{ \lambda \lambda^\prime } Q_{j \lambda^\prime}, 			
\end{align}

\begin{align}
  H_\epsilon &=  \frac{1}{2} \sum_i \left|  {\bm \pi}_i \right|^2 +
 \frac{1}{2} \sum_{i}  \left[ C_a \epsilon_{ia}^2 +  C_t \left( \epsilon_{it}^2 + \epsilon_{io}^2  \right)  + C_r  \left( \epsilon_{i 4}^2 + \epsilon_{i 5}^2 +  \epsilon_{i 6}^2  \right) \right ]  + P \sum_{i} \epsilon_{i a}, 
\end{align}

and
\begin{align}
  H_{Q \epsilon} &= g_a \sum_i \epsilon_{ia}  \left| { \bm Q }_i \right|^2 %  \nonumber \\
  -  g_t \sum_i \left[ \left( Q_{ix}^2 - Q_{iy}^2 \right) \epsilon_{io} + \frac{1}{\sqrt{3}} \left( 2 Q_{iz}^2 - Q_{iy}^2  - Q_{ix}^2 \right) \epsilon_{i t}   \right]  \nonumber \\
&- g_r \sum_i \left( Q_{ix} Q_{iy}  \epsilon_{i6} +  Q_{ix} Q_{iz}  \epsilon_{i5}  +  Q_{iy} Q_{iz}  \epsilon_{i4} \right).
\end{align}

Here, ${\bm \Pi}_i$ and  ${\bm \pi}_i$  are, respectively, the conjugate momenta of ${\bm Q}_i$ and ${\bm u}_i$. 
${\bm v}_{ij}^{\lambda \lambda^\prime}\, \left( \lambda, \lambda^\prime = x,y,z \right)$ is an interaction between the soft mode coordinates 
with Fourier transform  ${\bm v}_{{\bm R}+{\bm q}}^{\lambda \lambda^\prime} = v_{{\bm R}} \delta_{\lambda \lambda^\prime} + q^2 F_{\lambda \lambda^\prime}\left( \hat{\bm q} \right)$.
This form  is typical of cubic lattices with $F_{\lambda \lambda^\prime}\left( \hat{\bm q} \right)$  dependent on the direction of the unit wave-vector  $\hat{\bm q} = {\bm q} / q$ and independent of the  magnitude $q$.~\cite{SCowley1980a}
Within the local SCPA, the equations derived from the stationary property of the free energy
are independent of the  particular form of  $F_{\lambda \lambda^\prime}\left( \hat{\bm q} \right)$ as long as there is no self-interaction.~\cite{SPytte1972a}  $C_a = \left( C_{11} + 2 C_{12} \right)/3,$ is the bulk modulus, and $C_t = \left( C_{11} - C_{12} \right)/2,  C_r = C_{44}$, are deviatoric and shear moduli, respectively.
$g_a, g_t$,  and $g_r$ are coupling constants between the lattice and the strain degrees of freedom, and 
$P$ is an applied hydrostatic pressure. 
$\kappa_i$ is the lattice stiffness at site $i$; $\gamma_1$ and $\gamma_2$ 
are coefficients of the isotropic and anisotropic cubic anharmonicites, respectively. 

To account for quenched compositional disorder in mixed-compounds, we
note that at the mean field level the energy barriers between different lattice stuctures
depend on the ratio between the harmonic  and anharmonic coefficients of the model.~\cite{SCowley1980a} 
For simplicity, we thus consider a probability distribution $\mathcal{P}(\kappa_1,\kappa_2,...,\kappa_N)$ for the $\kappa$'s while assuming that the remaining parameters remain fixed. 

To solve the statistical mechanical problem posed by the Hamiltonian (\ref{eq:SHamiltonian}),
we use a variational formulation of a SCPA  in which the energies of the phonon excitations, displacement and strain order parameters are determined
from the minimization of the free energy.~\cite{SPytte1972a} 

\section{Statistical Mechanical Solution}

 We consider the trial probability distribution,
\begin{align}
	\label{Seq:rho_tr}
	\rho^{tr} = \frac{ e^{- \beta H^{tr}}}{Z^{tr}},
\end{align}
where $H^{tr}$ is the Hamiltonian of the local uncoupled problem,
\begin{align}
	H^{tr} = H_Q^{tr} +H_\epsilon^{tr},
\end{align}
\begin{align}
	 H_Q^{tr} &= \frac{1}{2} \sum_{i }\left| {\bm \Pi}_i \right|^2 + \frac{1}{2} \sum_{i,\alpha \beta} \left( Q_{i \alpha } - A_{i \alpha} \right)  \mathcal{M} _ {\alpha \beta} \left( Q_{i \beta } - A_{i \beta} \right),\\
	 H_\epsilon^{tr} &= \frac{1}{2} \sum_{i }\left| {\bm \pi}_i \right|^2+  \frac{1}{2} \sum_{i,\alpha \beta} \left( \epsilon_{i \alpha } - e_{i \alpha} \right) C_{\alpha \beta} \left( \epsilon_{i \beta } - e_{i \beta} \right).
\end{align}
$Z^{tr} = \mbox{Tr} e^{- \beta H^{tr}}$ is its normalization. $A_{i \alpha} =  \bra Q_{i \alpha} \ket $  and $e_{i \alpha} = \bra e_{i \alpha} \ket $
are the spontaneous displacement and strain order parameters which will be determined by minimization of the free energy; 
$ \mathcal{M}_{\alpha \beta}$ is the dynamical matrix with eigenfrequencies $\omega_\lambda$ of the non-ineracting problem ($v_{ij}^{\lambda \lambda^\prime}=0$).
Here, $ \bra ... \ket = \mbox{Tr} \left\{ 	\rho^{tr} .... \right\}$ denotes thermal average over the trial probability distribution (\ref{Seq:rho_tr}).
We set the long-range ordering associated with the condenstation of the $R_4^+$ mode at $R = (1,1,1)(\pi/a)$, 
by writing $A_{i \alpha} = A_{\alpha} e^{i {\bm R} \cdot {\bm r}_i }$, where ${\bm r}_i $ is the position vector the lattice site $i$. This corresponds to out-of-phase tilts
where $A_{i \lambda }$ changes sign from site to site.

\subsection{Free energy}
The free energy is calculated in the usual way $F = \bra H \ket + k_B T \bra  \ln \rho^{tr} \ket $,
\begin{align}
	 \label{Seq:free}
	F=F_Q+F_\epsilon + F_{Q \epsilon},
\end{align}
where,
\begin{align}
\frac{F_Q}{N} &= \frac{\ovl{\kappa}}{2} \bra\right.\left|{\bm Q}\right|^2\left.\ket
+
\frac{\gamma_1}{4}  \bra\right.\left|{\bm Q}\right|^4\left.\ket + \frac{\gamma_2}{2} \left( \bra Q_x^2 Q_y^2 \ket  + \bra Q_x^2 Q_z^2 \ket  +  \bra Q_y^2 Q_z^2 \ket  \right) 
-\frac{1}{2} \sum_{\lambda \lambda^\prime}  v_R^{\lambda \lambda^{\prime}} \bra Q_{\lambda} \ket \bra Q_{\lambda^\prime}  \ket  \nonumber \\ 
&- k_B T \sum_{\lambda} \left\{ \frac{\beta \omega_{\lambda} }{2} \coth \left(  \frac{\beta \omega_{\lambda} }{2}  \right) - \ln \left[  2 \sinh \left(  \frac{\beta \omega_{\lambda} }{2}  \right)  \right]\right\},
\end{align}
\begin{align}
\label{Seq:Feta}
\frac{F_{\epsilon}}{N} &=
\frac{1}{2} \left[  C_a  e_a^2  + C_t \left( e_t^2  + e_o^2  \right) + C_r \sum_{\nu =4}^6  e_\nu^2  \right]  + P e_a,
\end{align}
and,
\begin{align}
\label{Seq:FQeta}
\frac{F_{Q \epsilon}}{N} &= g_a e_a \bra\right.\left|{\bm Q}\right|^2\left.\ket 
-
g_t \left[ \left(   \bra Q_x^2 \ket -  \bra Q_y^2 \ket   \right) e_o 
 +  \frac{1}{\sqrt{3}} \left(   2 \bra Q_z^2 \ket -  \bra Q_x^2 \ket -  \bra Q_x^2 \ket  \right) e_t   \right] \nonumber \\
&-
g_r \left[ \bra Q_x Q_y \ket  e_6  +   \bra Q_x Q_z \ket  e_5  +   \bra Q_y Q_z \ket  e_4 \right],
\end{align}
with 
\begin{subequations}
	\label{Seq:averages}
	\begin{align}
	\bra Q_\lambda Q_{\lambda^\prime} \ket  &= A_{\lambda} A_{\lambda^\prime} + \psi_{\lambda \lambda^\prime},\\
	\bra Q_\lambda^2 Q_{\lambda^\prime}^2 \ket  &=  A_\lambda^2 A_{\lambda^\prime}^2+A_\lambda^2 \psi_{\lambda^\prime \lambda^\prime} + 4 A_\lambda A_{\lambda^\prime} \psi_{\lambda \lambda^\prime} + A_{\lambda^\prime}^2 \psi_{\lambda \lambda} + \psi_{\lambda \lambda} \psi_{\lambda^\prime \lambda^\prime} + 2 \psi_{\lambda \lambda^\prime}^2.
	\end{align}
\end{subequations}
$\psi_{\lambda \lambda^\prime}$ are the local OP fluctuations given as follows,
\begin{align}
		\psi_{\lambda \lambda^\prime} 
	=\sum_{\nu} b_{\lambda \nu }^\dagger  b_{\nu \lambda^\prime } \left( \frac{1}{2 \omega_\nu } \coth \left( \frac{\beta \omega_\nu }{2} \right) \right),
\end{align}
where $b_{\nu \lambda^\prime } $ is an unitary transformation that diagonalizes $\mathcal{M} _ {\alpha \beta}$.
$\overline{\kappa}$ the lattice stiffness averaged over compositional disorder.
In writing Eq.~(\ref{Seq:free}), we have ignored all terms that do not depend on $A_{\lambda}, e_\alpha$, and $\omega_{\lambda}$ as they do not
have an effect on the miminization procedure.

Minimization  of the  free energy (\ref{Seq:free}) with respect to the strains $e_\alpha$ gives the following result,
\begin{subequations}
	\label{Seq:es}
	\begin{align} 
		e_a &=  -\frac{g_a}{C_a} \bra\right.\left|{\bm Q}\right|^2\left.\ket  - \frac{P}{C_a},  \\
		e_t &=  \frac{g_t}{\sqrt{3} C_t}  \left( 2 \bra Q_z^2 \ket  - \bra Q_x^2 \ket -  \bra Q_y^2 \ket \right),  \\
		e_o &= \frac{g_t}{C_t}   \left( \bra Q_x^2 \ket  -  \bra Q_y^2 \ket  \right),  \\
		e_4 &=   \frac{g_r}{C_r}  \bra Q_y Q_z \ket, ~~
		e_5 =   \frac{g_r}{C_r}  \bra Q_x Q_z \ket, ~~ 
		e_6 =   \frac{g_r}{C_r}  \bra Q_x Q_y \ket.
	 \end{align}
\end{subequations}

By substituting the strains of Eq.~(\ref{Seq:es}) into the Eq.~(\ref{Seq:free}),
we obtain a free energy which  depends on the displacements ${\bm Q}$  only,
where,
\begin{align}
	\label{Seq:free_renormalized}
	\frac{\tilde{F}}{N} &=  \frac{\tilde{\kappa}}{2} \bra\right.\left|{\bm Q}\right|^2\left.\ket
+
\frac{\gamma_1}{4}  \bra\right.\left|{\bm Q}\right|^4\left.\ket + \frac{\gamma_2}{2} \left( \bra Q_x^2 Q_y^2 \ket  + \bra Q_x^2 Q_z^2 \ket  +  \bra Q_y^2 Q_z^2 \ket  \right) 
-\frac{1}{2} \sum_{\lambda \lambda^\prime}  v_R^{\lambda \lambda^{\prime}} \bra Q_{\lambda} Q_{\lambda^\prime}  \ket 
 \nonumber, \\
 &- k_B T \sum_{\lambda} \left\{ \frac{\beta \omega_{\lambda} }{2} \coth \left(  \frac{\beta \omega_{\lambda} }{2}  \right) - \ln \left[  2 \sinh \left(  \frac{\beta \omega_{\lambda} }{2}  \right)  \right]\right\}, \nonumber \\
  &-\left( \frac{3}{2} \frac{g_a^2}{C_a} + \frac{2}{3}\frac{g_t^2}{C_t} \right)  \left( \bra Q_x^2 \ket  +  \bra Q_y^2 \ket  +  \bra Q_z^2 \ket  \right)^2  \nonumber  \\
								 &+ \frac{2g_t^2}{C_t} \left(  \bra Q_x^2 \ket  \bra Q_y^2 \ket  + \bra Q_x^2 \ket  \bra Q_z^2 \ket + \bra Q_y^2 \ket  \bra Q_z^2 \ket  \right) \nonumber \\
								  &-\frac{1}{2}\frac{g_r^2}{C_r} \left( \bra Q_x Q_y \ket^2 + \bra Q_x  Q_z \ket^2 + \bra Q_y Q_z \ket^2  \right) - \frac{P^2}{2C_a},
\end{align}
where $\tilde{\kappa} \equiv \ovl{\kappa} - 2 g_a P / C_a $.  The free energies (\ref{Seq:free}) \& (\ref{Seq:free_renormalized}) are the starting point for our calculation of the thermodynamic quantities of interest. 

\newpage

\subsection{Soft Mode Frequencies}

The soft mode frequencies are computed from
the free energy~(\ref{Seq:free}) with the $e_{\alpha}$ constant 
and then  must be evaluated at the equilibrium points
given in Eq~(\ref{Seq:es}).  
This is because the frequency of the  acoustic modes associated with 
uniform strains vanishes in the long-wavelength limit.~\cite{SSlonczewski1970a}

The dynamical matrix $ \mathcal{D}_{\lambda \lambda^\prime}$ of the interacting problem is calculated from the free energy~(\ref{Seq:free}),
\begin{align*}
 \mathcal{D}_{xx} = \frac{\partial^2 \bra H  \ket}{\partial A_x \partial A_x} &= \kappa + \left( 2 \gamma_1 - \gamma_2 \right)   \bra Q_x^2 \ket  + \left( \gamma_1 + \gamma_2 \right) \bra\right.\left|{\bm Q} \right|^2\left.\ket      \\ 
&~~~~~~~~~~~~~~~~~~~~~~~~~~~~~~~
 + 2 g_a e_a - 2 g_t \left( e_o - \frac{e_t}{\sqrt{3}}  \right)  -  v_R,   \\
 \mathcal{D}_{yy} = \frac{\partial^2 \bra H  \ket}{\partial A_y \partial A_y} &= \kappa + \left( 2 \gamma_1 - \gamma_2 \right)    \bra Q_y^2 \ket   + \left( \gamma_1 + \gamma_2 \right)  \bra\right.\left|{\bm Q} \right|^2\left.\ket      \\ 
&~~~~~~~~~~~~~~~~~~~~~~~~~~~~~~~
 + 2 g_a e_a + 2 g_t \left( e_o + \frac{e_t}{\sqrt{3}}  \right) -  v_R, \\
  \mathcal{D}_{zy} = \frac{\partial^2 \bra H  \ket}{\partial A_z \partial A_z} &= \kappa + \left( 2 \gamma_1- \gamma_2 \right)   \bra Q_z^2 \ket  + \left( \gamma_1 + \gamma_2 \right) \bra\right.\left|{\bm Q} \right|^2\left.\ket    \\
&~~~~~~~~~~~~~~~~~~~~~~~~~~~~~~~
 + 2 g_a e_a - 4  g_t \frac{e_t}{\sqrt{3}}  -   v_R, \\
 \mathcal{D}_{xy} = \frac{\partial^2 \bra H  \ket}{\partial A_x \partial A_y} &=  2 \left( \gamma_1 + \gamma_2 \right)  \bra Q_x Q_y \ket    - g_r e_6, \\
  \mathcal{D}_{xz} = \frac{\partial^2 \bra H  \ket}{\partial A_x \partial A_z} &= 2 \left( \gamma_1 + \gamma_2 \right) \bra Q_x Q_z \ket   - g_r e_5, \\
 \mathcal{D}_{yz} = \frac{\partial^2 \bra H  \ket}{\partial A_y \partial A_z} &=  2 \left( \gamma_1+ \gamma_2 \right) \bra Q_y Q_z \ket  - g_r e_4,
\end{align*}
where $ \bra Q_{\lambda} Q_{\lambda^\prime} \ket$ is given by Eq.~(\ref{Seq:averages}). To proceed further, we now consider the cubic and rhombohedral phases separately. 

\subsection{Cubic phase }

In the cubic phase, $ A_x=A_y=A_z=0,  \bra Q_x^2 \ket  =  \bra Q_y^2 \ket =  \bra Q_z^2 \ket = \psi_0, \bra Q_x Q_y \ket =  \bra Q_x Q_z \ket =  \bra Q_y Q_z \ket  =0,  e_a \neq 0,  e_t = e_o= e_4=e_5=e_6=0, b_{\lambda \lambda^\prime} = \delta_{\lambda \lambda^\prime}$. Thus, the dynamical matrix is given as follows,
\begin{align*}
 \mathcal{D}_{xx}=\mathcal{D}_{yy}=\mathcal{D}_{zz} &= \overline{\kappa} + \left( 5 \gamma_1 + 2 \gamma_2  \right)  \psi_0 + 2 g_a   e_a   - v_R,\\
 \mathcal{D}_{xy}= \mathcal{D}_{xz}=\mathcal{D}_{yz} &=0.
\end{align*}
The diagonalization of $ \mathcal{D} _{ \lambda \lambda^\prime} $ gives
a triply degenerate zone-boundary soft mode frequency,
\begin{align}
	\label{Seq:softmode}
%	\Omega_{{\bm q}_0}^2  &= - \omega_{{\bm q}_0}^2  + \left( 5 \gamma_1 + 2 \gamma_2 \right)  \psi_0 +  2 g_a e_a.
		\Omega_{R_4^+}^2  &= \omega_{R_4^+}^2  + \left( 5 \gamma_1 + 2 \gamma_2 \right)  \psi_0 +  2 g_a e_a.
\end{align}
Here,  $\omega_{R_4^+} \equiv \sqrt{\overline{\kappa} - v_R}$ is the frequency of a purely harmonic model.
$\psi_0$ are the local OP fluctuations in the cubic phase,
\begin{align}
	\label{Seq:psi0}
	\psi_0 =  \frac{ 1 }{ 2  \omega_R }  \coth\left( \frac{ \beta  \omega_{R} }{2} \right),
\end{align}
with $\omega_{R} = \sqrt{\Omega_{R_4^+}^2 + v_R}$. 
The change in volume is given by the volumetric strain,
\begin{align}
	\label{Seq:NTE}
	\frac{\Delta V}{V_0} = e_a =  - \frac{3g_a}{C_a}  \psi_0 - \frac{P}{C_a},
\end{align}
where $V_0$ is a reference volume. 

We now calculate the  Gr\"{u}neisen parameter  associated with $R_4^+$. 
From Eq.~(\ref{Seq:softmode}), we find that the temperature and pressure dependence of $\gamma_{R_4^+}$ 
is entirely determined by the phonon energy,~\cite{SVolker82a}
\begin{align}
	\label{Seq:gruneisen_c}
	\gamma_{R_4^+} = - \frac{\partial \ln \Omega_{R_4^+}}{\partial e_a} = - \frac{g_a}{\Omega_{R_4^+}^2}.
\end{align}

 Equations~(\ref{Seq:softmode})-(\ref{Seq:NTE}), determine self-consistently the temperature and pressure dependence of 
$\Omega_{R_4^+} $, and $e_a$. 

\subsection{Rhombohedral phase}

In the r-phase, $A_x=A_y=A_z=A/\sqrt{3}, \psi_1 \equiv \psi_{xx}=\psi_{yy}=\psi_{zz}, \psi_4 \equiv \psi_{xy}=\psi_{xz}=\psi_{yz}, e_a \neq 0,  e_o=e_t=0, e_r \neq 0$ and, 
\begin{align}
	b_{\lambda \lambda^\prime} 
	= \begin{pmatrix}
		\frac{1}{\sqrt{6}} & \frac{1}{\sqrt{2}} & \frac{1}{\sqrt{3}} \\		
		\frac{1}{\sqrt{6}} & -\frac{1}{\sqrt{2}} & \frac{1}{\sqrt{3}} \\		
		-\frac{2}{\sqrt{6}} & 0 & \frac{1}{\sqrt{3}} \\		
	\end{pmatrix}.
\end{align}
Thus, the dynamical matrix is given as follows,
\begin{align*}
	\mathcal{D}_{xx}=\mathcal{D}_{yy}=\mathcal{D}_{zz} &= \overline{\kappa} + \left( 2 \gamma_1 - \gamma_2 \right)  \left( \frac{A^2}{3} + \psi_1 \right) + \left( \gamma_1 + \gamma_2  \right) \left( A + 3 \psi_1 \right)  + 2 g_a e_a - v_R,\\
	\mathcal{D}_{xy}= \mathcal{D}_{xz}=\mathcal{D}_{yz} &=  2 \left( \gamma_1 +\gamma_2 \right)  \left(\frac{A^2}{3}  + \psi_4  \right) - g_r e_r .
\end{align*}

The diagonalization of $\mathcal{D}_{\lambda \lambda}$ together with the minimization of the free energy~(\ref{Seq:free_renormalized}) with respect to $A$
gives the following result,
\begin{subequations}
	\label{Seq:M3+}
\begin{align}
	\Omega_{E_g}^2 &= \omega_{R_4^+}^2 + \left( 5 \gamma_1 +2 \gamma_2 \right)  \left( \frac{A^2}{3}  + \psi_1 \right)   +2 g_a e_a - 2 \left( \gamma_1 +\gamma_2 \right)  \left(\frac{A^2}{3}  +\psi_4  \right)  + g_re_r, \\
		\Omega_{A_{1g}}^2 &=\omega_{R_4^+}^2 + \left( 5 \gamma_1 +2 \gamma_2 \right)  \left( \frac{A^2}{3}  + \psi_1 \right)   + 2 g_a e_a + 4 \left( \gamma_1 +\gamma_2 \right)  \left(\frac{A^2}{3}  +\psi_4  \right) -  2 g_r e_r, \\
		\Omega_{A_{1g}}^2  &= 2 \left( 3 \gamma_1 + 2 \gamma_2 \right) \frac{A^2}{3},
\end{align}
\end{subequations}
$e_a$ and $e_r$ are volume and shear strains, respectively
\begin{subequations}
	\label{Seq:strains_r}
\begin{align}
	\label{Seq:ear}
	e_a &= - \frac{3g_a}{C_a}  \left( \frac{A^2}{3} +  \psi_1 \right) - \frac{P}{C_a},\\
	\label{Seq:er}
	e_r &=  \frac{g_r}{C_r} \left( \frac{A^2}{3} +  \psi_4 \right).
\end{align}
\end{subequations}
$\psi_1$ and $\psi_4$ are fluctuations of the OP in the r-phase,
\begin{subequations}
	\label{Seq:psi1_psi4}
\begin{align}
	\psi_1 &= \frac{1}{3} \left( \frac{1}{2 \omega_{A_{1g}} } \coth \left( \frac{\beta \omega_{A_{1g}} }{2} \right) + \frac{1}{\omega_{E_g} } \coth \left( \frac{\beta \omega_{E_g} }{2} \right)   \right),\\
	\psi_4 &= \frac{1}{3} \left( \frac{1}{2 \omega_{A_{1g}} } \coth \left( \frac{\beta \omega_{A_{1g}} }{2} \right) - \frac{1}{2\omega_{E_g} } \coth \left( \frac{\beta \omega_{E_g} }{2} \right)   \right),  
\end{align}
\end{subequations}
where $\omega_{E_g,A_{1g}}=\sqrt{\Omega_{{E_g,A_{1g}}}^2 + v_R}$.

As mentioned above, the temperature and pressure dependence of the the Gr\"{u}neisen parameters of the $E_g$ and $A_{1g}$ phonons 
are again entirely determined by their corresponding energies,
\begin{align}
	\gamma_{E_g,A_{1g}} =  - \frac{\partial \ln \Omega_{E_g,A_{1g}}}{\partial e_a} = - \frac{g_a}{\Omega_{E_g,A_{1g}}^2}.
\end{align}

 Equations~(\ref{Seq:M3+})-(\ref{Seq:psi1_psi4}), determine self-consistently the temperature and pressure dependence of 
$\Omega_{E_g/A_{1g}}$, $A$, $e_r$ and $e_a$. 

\subsubsection{Classical Limit}

It is useful to consider the classical limit of the above results, as it allow us to 
derive analytical expressions for several relevant macroscopic quantities in terms
of microscopic parameters. We consider the high-$T$ cubic phase. 

In the classical limit ($\beta \omega \ll 1 $) and near the structural transition, $\psi_0\simeq  k_B T /v_R$, thus,
\begin{align}
	\label{Seq:alphaV}
	\frac{\Delta V}{V_0} \simeq  \alpha_V T - \frac{P}{C_a}, ~~~~\alpha_V = - \frac{3g_a k_B}{C_a  v_R} 
\end{align}
where $\alpha_V$ is the coefficient of thermal expansion (CTE). 

\begin{figure}[htp!]
	\centering
			\includegraphics[scale=0.45]{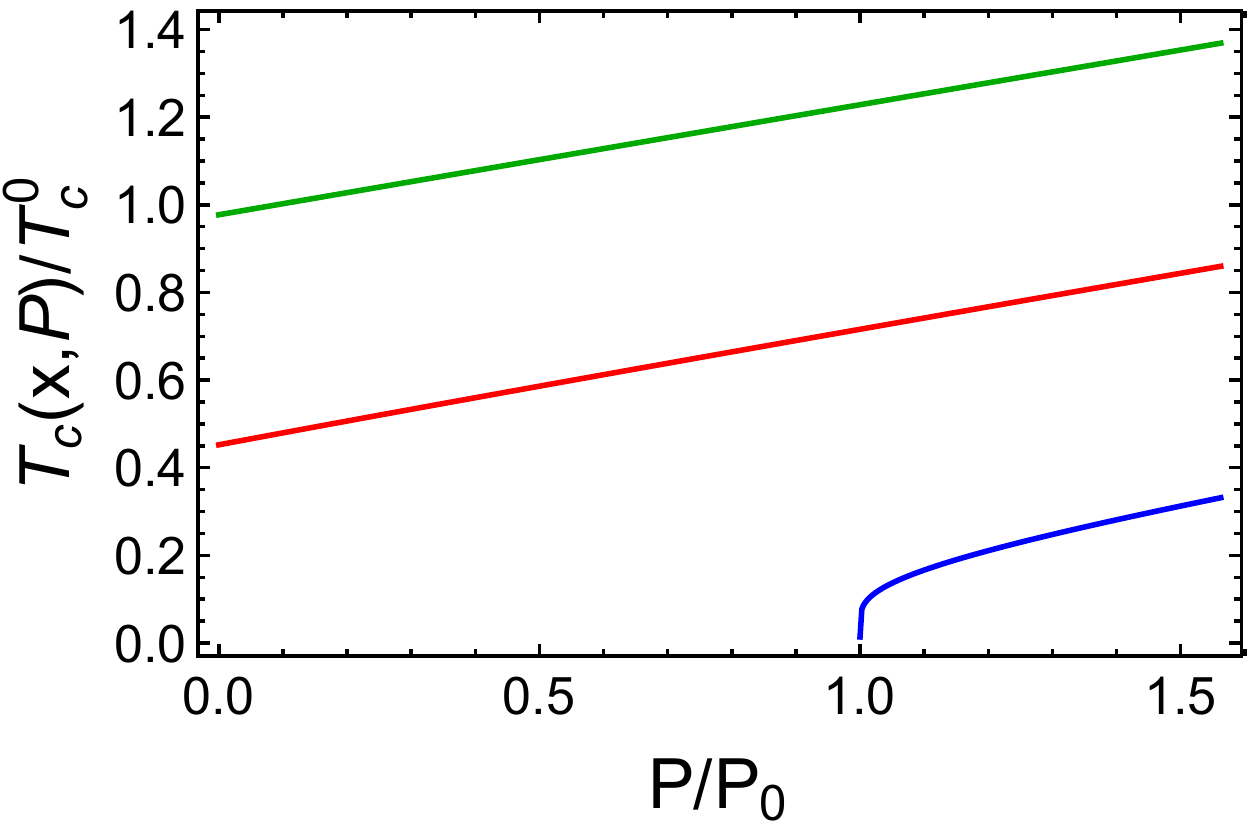}
		\includegraphics[scale=0.45]{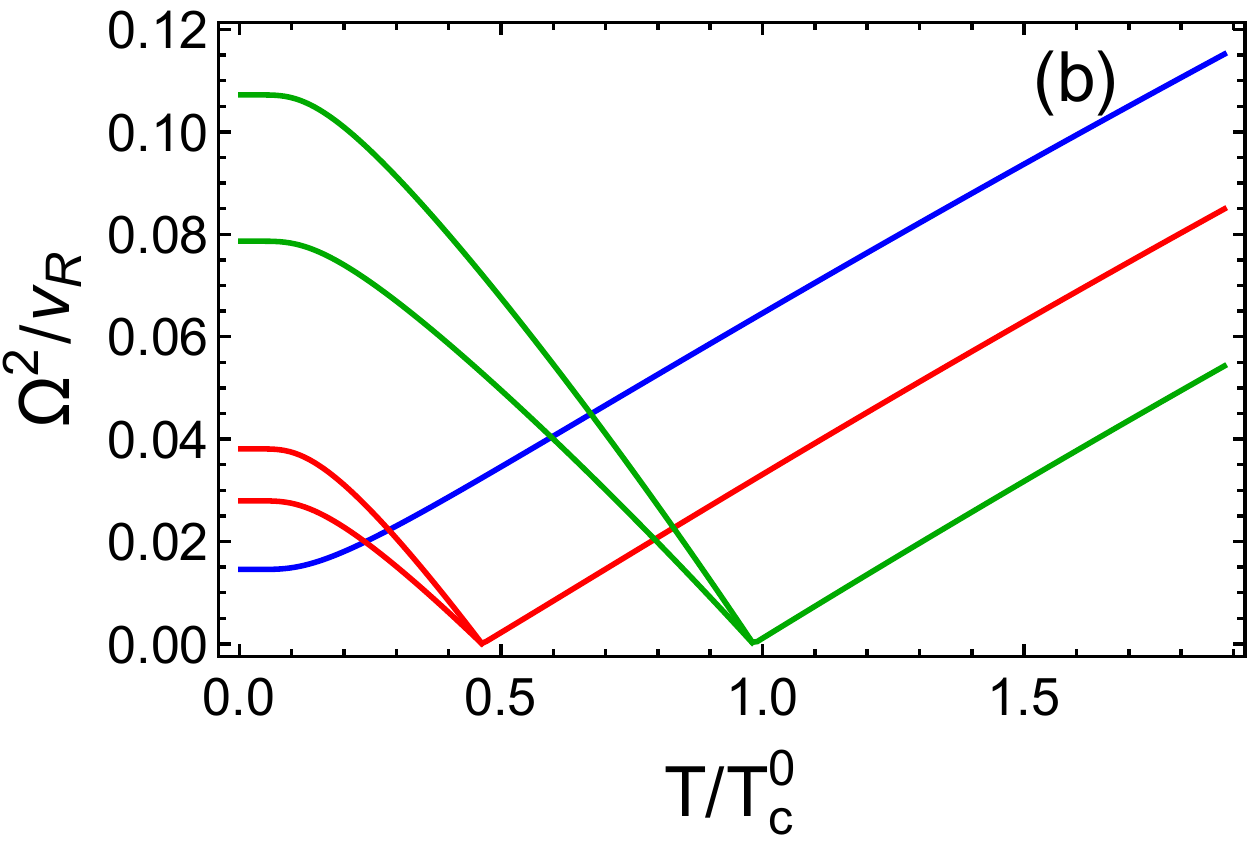}
				\includegraphics[scale=0.45]{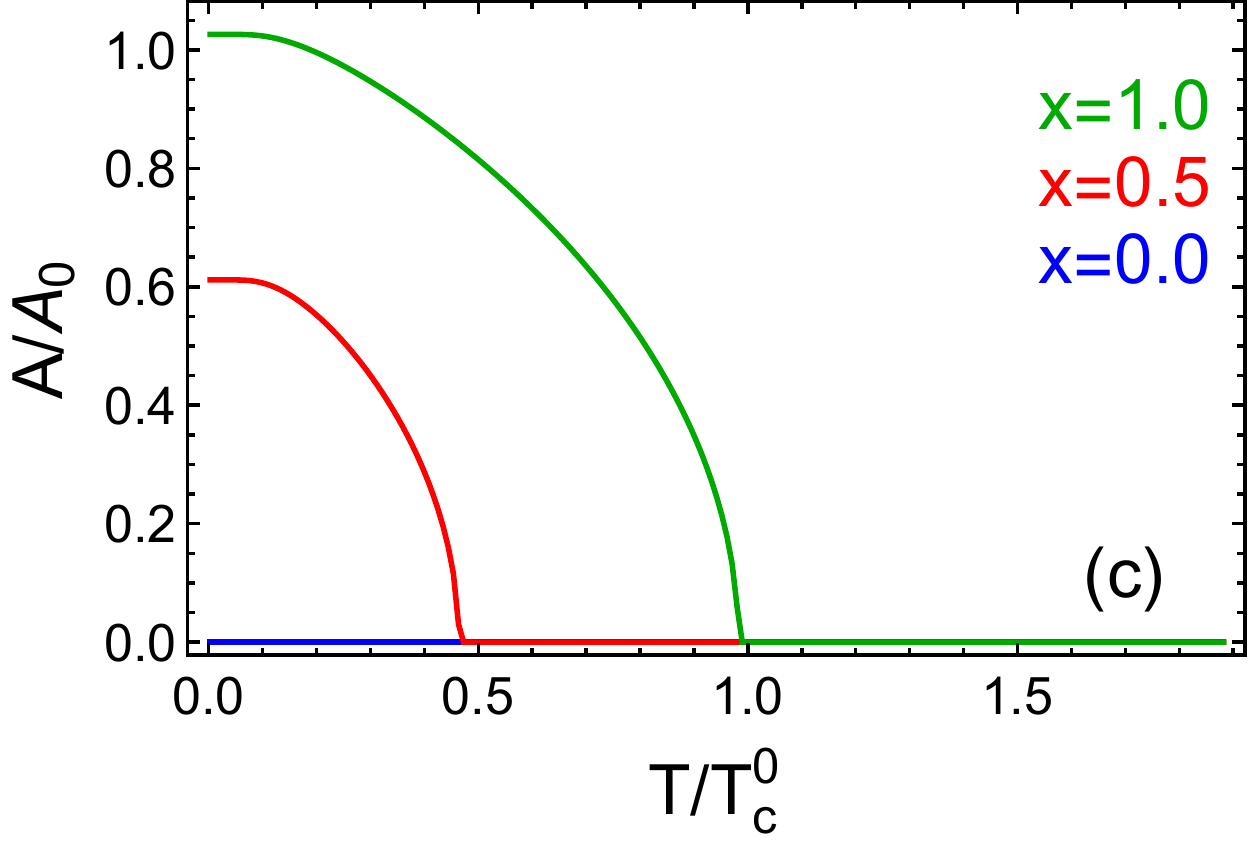}
	\caption{Calculated (a) phase diagram (b) soft mode frequencies and (c) order parameter for Sc$_x$Ti$_{1-x}$F$_3$.}
	\label{Sfig:phase_transition}
\end{figure}

We find that he $T$ and $P$ dependence of $\Omega_{R_4^+}$ matches that of Landau theory, as expected,~\cite{SPytte1972a}
\begin{align}
	\Omega_{R_4^+}^2 = -\omega_{R_4^+}^2 t,
\end{align}
where  $t = \frac{T-T_c(x,P)}{T_c(x,P)}$ is a reduced temperature and 
$T_c(x,P)$ is a pressure dependent transition temperature given as follows,
\begin{align}
		\label{Seq:TcP}
		T_c(x,P) = T_c(x,0) +   \frac{2 g_a / C_a}{\gamma k_B/v_R}   P,
\end{align}
where $T_c(x,0)$ is the transiton temperature at ambient pressure,
\begin{align}
	\label{Seq:Tc}
	T_c(x,0) =   \frac{-\omega_{R_4^+}^2}{ \gamma k_B/v_R },
\end{align}
with $\gamma \equiv 5 \gamma_1 + 2 \gamma_2 -6 g_a^2/C_a$. Note that when there is a c-r transition, the high symmetry phase is unstable in the purely harmonic approximations and thus $\omega_{R_4^+}^2 < 0$. The $x$ dependence in $T_c \left( x,P \right)$ is through $\omega_{R_4^+}$, which in turn depends $\overline{\kappa}$, i.e., the lattice stiffness averaged over compositional disorder.  Note that for $g_a>0$, hydrostatic pressure destabilizes the c-phase.
From Eq.~(\ref{Seq:TcP}), we find a proportionality relation between the slope of the $T-P$ phase diagram and the CTE,
\begin{align}
		\label{Seq:dTcdP}
		\frac{dT_c}{dP} =  \frac{2 g_a / C_a}{ \gamma k_B/v_R }  \propto -\alpha_V.
\end{align}

We conclude the presentation of our model here. 

\section{Model parameters}

The model parameters are given in Table~\ref{t:parameters} and were obtained from fits
to experiments in Sc$_x$Ti$_{1-x}$F$_3$.~\cite{SHandunkanda2015a, SMorelock2014a} We have assumed independent
bimodal distributions for the stiffnesses $\mathcal{P}(\kappa_1, \kappa_2,...,\kappa_N)=\prod_{i=1}^N \mathcal{P}(\kappa_i)$, with $\mathcal{P}(\kappa_i) = x \delta(\kappa_i - \kappa_{\text{Ti}}) + \left(1- x\right)\delta(\kappa_i - \kappa_{\text{Sc}})$ and
 $\kappa_{\text{Ti}/\text{Sc}}$ are the lattice stiffnesses of the pure compounds. 
With this choice, $\overline{\kappa} = x \kappa_{\text{Ti}} + \left(1- x\right) \kappa_{\text{Sc}} $.
The resulting $T-P$ phase diagram, phonon frequencies and order parameter for several compositions are shown in Fig.~\ref{Sfig:phase_transition}.

	\begin{table}[h!]
		\centering
		\caption{Model parameters for Sc$_x$Ti$_{1-x}$F$_3$.} 
		\label{t:parameters}
		\begin{tabular}{rc} \hline \hline
			$\kappa_{\text{Ti}}\,$[meV$^{2}$] & $161$    \\
						$\kappa_{\text{Sc}}\,$[meV$^{2}$] & $ 173$    \\
           	$ v_R\,$[meV$^2$] & $173\,$   \\          
           	$ \gamma_1\,$[meV$^3$] & $21.8\,$   \\          
           	$ \gamma_2\,$[meV$^3$] & $-19.4\,$   \\          

           	$ g_a\,$[meV$^2$] & $0.023\,$   \\
		    	$ g_r\,$[meV$^2$] & $0.019\,$   \\ 
		    	$  C_a\,$[meV] & $1.0\,$ \\        
  			$  C_r\,$[meV] & $0.22\,$ \\         \hline 
    		\end{tabular}
	\end{table}

\end{document}